\documentclass[preprint, letterpaper, 12pt]{aastex}
\shorttitle{Modeling the Envelope Around L1157}
\shortauthors{Chiang et al.}

\begin{document}
\title{
The Envelope and Embedded Disk around the Class 0 Protostar L1157-mm:  
Dual-wavelength Interferometric Observations and Modeling 
\footnote{Accepted for publication in ApJ}
}
\author{Hsin-Fang Chiang\altaffilmark{2}\altaffilmark{3}, 
Leslie W. Looney\altaffilmark{2}, 
John J. Tobin\altaffilmark{4}  
} 
\altaffiltext{2}{Department of Astronomy, University of Illinois at
Urbana-Champaign, 1002 West Green Street, Urbana, IL 61801
}
\altaffiltext{3}{Institute for Astronomy, University of Hawaii at Manoa,
 640 North Aohoku Place, Hilo, HI 96720; hchiang@ifa.hawaii.edu }
\altaffiltext{4}{Hubble Fellow, National Radio Astronomy Observatory, Charlottesville, Virginia}

\begin{abstract}

We present dual-wavelength observations   and modeling 
of the nearly edge-on Class 0 young stellar object L1157-mm.  
Using 
the Combined Array for Research in Millimeter-wave Astronomy,   
a nearly spherical 
structure is seen from the circumstellar envelope at the 
size scale of 10$^2$ to 10$^3$ AU in both 1 mm and 3 mm dust emission.  
Radiative transfer modeling is performed to compare data 
with theoretical envelope models, including a power-law envelope 
model and the Terebey-Shu-Cassen model. 
Bayesian inference is applied for parameter estimation 
and information criteria is used for model selection.  
%
The results prefer the power-law envelope model 
against the  Terebey-Shu-Cassen model. 
%
In particular, for the power-law envelope model, 
a steep density profile with an index of $\sim$2 is inferred. 
Moreover, 
the dust opacity spectral index $\beta$ is estimated to be 
$\sim$0.9, implying that 
grain growth has started at L1157-mm. 
Also, the unresolved disk component is constrained    
to be $\lesssim$ 40 AU in radius and 
$\lesssim$ 4-25 M$_{Jup}$ in mass. 
However, the estimate of the embedded disk component 
relies on the assumed envelope model.  


\end{abstract}

\section{Introduction }


Protostars are surrounded by their natal envelopes  
in the earliest stage of evolution. 
These envelopes supply the material that is 
actively infalling onto the embedded star-disk system,   
and their properties can affect the subsequent evolution. 
Much theoretical work has  been done to address the 
collapse process and the physical properties of the envelope, 
varying from self-similar solutions 
to numerical calculations including 
rotation and magnetic fields  
\citep[e.g., ][]{Larson1969,Penston1969,Hunter1977,Shu1977,TSC1984,Whitworth1985,Galli1993,MasunagaInutsuka2000,Tassis2005a,Hennebelle2008}.    
While various theoretical models of the collapsing envelope 
have been suggested,
distinguishing between the suggested theoretical models has always been an observational challenge.  
For example, the model of \citet*[][hereafter the TSC model]{TSC1984} 
has been widely used 
and consistent results have been obtained, 
especially in fitting the spectral energy distributions of 
unresolved young stellar objects  
\citep[e.g.,][]{Robitaille2007}, 
but it is based on the Shu (1977) model, which 
could not fit a sample of Class 0 protostars with reasonable ages.  
(Looney et al. 2003). 






To better constrain the envelope structure, 
we carry out complete modeling with dual-wavelength millimeter data.  
At such wavelengths, the continuum is dominated by dust emission    
from the envelope and the embedded disk. 
Interferometry is a useful tool to probe the structure of protostellar 
envelopes as it measures emission at various spatial scales, 
leading to a more complete analysis for the envelope.  
By comparing the predicted envelope
structure with interferometric observations, 
theoretical collapse models can be tested  
\citep[e.g.,][]{Chiang2008,Maury2010}. 
%
A better understanding of the envelope also enables us to constrain the 
physical properties of the embedded disk component. 



In this paper, we focus on the edge-on Class 0 protostar 
L1157-mm (also known as L1157-IRS or IRAS 20386+6751).  
The distance to L1157-mm is around 200-450 pc \citep[][]{Straizys1992,Kun1998,Kun2008}; 
here we follow \citet{Looney2007} and adopt 250 pc.  
A chemically active outflow driven by L1157-mm has been 
detected in multiple species 
\citep{Gueth1996,Bachiller2001,Nisini2010}.  
Perpendicular to the outflow orientation,  
a flattened envelope with a linear size of $\sim$20,000 AU  
is seen in 8 $\mu$m absorption,  
N$_2$H$^+$ emission, and NH$_3$ emission,  
showing complex kinematics from 
rotation, infall, and outflow in the envelope 
\citep{Looney2007,Chiang2010,Tobin2011}. 
On the other hand, dust continuum traces the envelope structure  
as well as  reveals a compact core 
\citep[e.g.,][]{Gueth2003}.   
Additionally, the presence of a circumstellar disk embedded inside the envelope is  
suggested by methanol observations
 \citep[][]{Goldsmith1999L1157,Velusamy2002}.

We have collected 1 mm and 3 mm interferometric data 
at multiple array configurations using 
the Combined Array for Research in Millimeter-wave Astronomy 
\citep[CARMA;][]{Woody2004}   
\footnote{http://www.mmarray.org/}.  
\S \ref{sec:obs}  gives an overview of the observations and the data reduction.  
Details of the modeling are addressed in \S \ref{sec:modeling} 
with a further supplement on our statistical approach in the Appendix. 
The results are presented in \S \ref{sec:result}, their  
implications are discussed in \S \ref{sec:discussion}, 
and a summary is given in \S \ref{sec:summary}.  



\section{Observations and Data Reduction} \label{sec:obs}

L1157-mm was observed by the 15-element CARMA  
between Oct 2007 and  Jan 2010.   
At that time, the science array of CARMA 
consisted of six 10.4-meter antennas 
and nine 6.1-meter antennas.  
Dust continuum at both 1 mm and 3 mm bands was  observed 
using multiple array configurations, 
as summarized in Table \ref{tabObs}. 

\begin{deluxetable}{cclccccc}
\tabletypesize{\footnotesize}
\tablecolumns{8}
\tablewidth{0pc}
\tablecaption{Summary of Observations}
\tablehead{
\colhead{Frequency}         &
\colhead{Array }    &
\colhead{Date }    &
\colhead{Observing }    &
\colhead{Bandpass}    &
\colhead{Flux}    &
\colhead{Beam Size \tablenotemark{a} } &
\colhead{Beam P.A. \tablenotemark{a} }  
\\
\colhead{(GHz)}      &
\colhead{Config.}  & 
\colhead{}  & 
\colhead{Time (hr)}  & 
\colhead{Calibrator }    &
\colhead{Calibrator }    &
\colhead{($\arcsec$) }  & 
\colhead{(degree)}
\\
\colhead{(1)}  & 
\colhead{(2)}  & 
\colhead{(3)}  & 
\colhead{(4)}  & 
\colhead{(5)}  & 
\colhead{(6)}  & 
\colhead{(7)}  & 
\colhead{(8)}    
}
\startdata
229 & B & 2007-12-17 \tablenotemark{b}
        & 1.5  
        & 3C454.3  &  MWC349 
        & 0.4$\times$0.3  & -75 \\
    & C & 2008-04-13        
        & 3.1 
        & 3C454.3  &  MWC349
        & 1.0$\times$0.8  & -66 \\
    & D & 2008-03-07        
        & 3.1 
        & 3C454.3  &  MWC349
        & 2.3$\times$2.0  & -29 \\ 
91  & A & 2009-01-27  \tablenotemark{bc}
        & 3.3 
        & 1642+689 & MWC349 
        & 0.4$\times$0.3  & -65    \\ 
    &   & 2010-01-26  \tablenotemark{bd}
        & 2.2 
        & 3C273 & MWC349
        &  &  \\ 
    &   & 2010-02-01  \tablenotemark{bd}
        & 2.9 
        & 3C454.3  & Neptune 
        &  &  \\ 
    & B & 2007-11-17   \tablenotemark{b}     
        & 4.5 
        & 1751+096 & MWC349  
        & 0.9$\times$0.8   & -83    \\ 
    &   & 2007-11-19   \tablenotemark{b} 
        & 2.4 
        & 3C454.3  & Neptune 
        &  &  \\
    &   & 2007-11-20   \tablenotemark{b} 
        & 1.5 
        & 3C273    & 3C273   
        &  &  \\
    &   & 2009-12-14   \tablenotemark{b} 
        & 0.6 
        & 3C454.3  & Uranus 
        &  &  \\
    &   & 2009-12-15    \tablenotemark{bd}
        & 5.4 
        & 3C345    & MWC349
        &  &  \\
    & C & 2007-10-03        
        & 2.5 
        & 1751+096 & MWC349    
        & 2.3$\times$2.0   & -87    \\
    &   & 2007-10-05        
        & 2.8 
        & 1751+096 & MWC349    
        &  &  \\
    &   & 2008-04-05        
        & 4.8 
        & 3C273    & MWC349    
        &  &  \\
    & D & 2008-02-29        
        & 3.1 
        & 3C454.3  & Uranus 
        & 6.0$\times$5.2   & 84    \\
    &   & 2009-03-19 
        & 6.0 
        & 3C345    & MWC349    
        &  &  \\
    &   & 2009-03-20 
        & 6.2 
        & 3C345    & MWC349    
        &  &  \\
    &   & 2009-03-27 
        & 0.8 
        & 1642+689 & MWC349    
        &  &  \\
    &   & 2009-03-29 
        & 3.0
        & 1642+689 & MWC349    
        &  &  \\
    & E & 2008-10-02        
        & 4.2 
        & 3C454.3 & Uranus 
        & 11.5$\times$10.2   & 78 \\ 
    &   & 2008-10-05 
        & 3.5 
        & 3C84     & MWC349    
        &  &  \\
\enddata
\tablenotetext{a}{
The synthesized beam of the combined data with natural weighting 
at each array configuration 
}
\tablenotetext{b}{
Track examined with the secondary phase calibrator 2009+724
}
\tablenotetext{c}{
Track using the primary phase calibrator 1849+670  
instead of 1927+739. 
}
\tablenotetext{d}{
Track calibrated with C-PACS 
using the atmospheric calibrator 2022+616
}
\label{tabObs}
\end{deluxetable}

The phase center of the observations before September 2008 was 
$\alpha$ = 20$^h$39$^m$06\fs20,   
$\delta$ = 68\degr02\arcmin15\farcs9, 
and shifted to 
$\alpha$ = 20$^h$39$^m$06\fs26,   
$\delta$ = 68\degr02\arcmin15\farcs8 
afterwards as more precise coordinates were determined 
by high resolution observations. 
However, all data presented here have been corrected to  
have the common phase center at 
$\alpha$ = 20$^h$39$^m$06\fs26,   
$\delta$ = 68\degr02\arcmin15\farcs8 (J2000).  

The main phase calibrator for all tracks was 1927+739  
(with the exception of one A-array track) 
and was observed with a phase calibrator-source cycle of 10-15 minutes.  
For all A- and B-array observations, a weaker quasar, 2009+724, 
was observed as the secondary phase calibrator.   
The secondary phase calibrator was not used in the calibration process; 
instead, it is used to check the point source response.  
For 3 mm A- and B-array tracks observed in winter 2009-2010, 
the CARMA Paired Antenna Calibration System \citep[C-PACS;][]{PACS}
was employed to calibrate the atmospheric phase variation on short time-scale. 
With C-PACS, a reference array continuously monitors 
a nearby quasar, called the atmospheric calibrator, for atmospheric delay,  
while the science array observes the science target.  
The eight 3.5-meter antennas, from the previous Sunyaev-Zel'dovich Array (SZA), 
were used as the reference array.   
The C-PACS correction is effective for data with long baseline. 
%

The data reduction, calibration, and imaging 
were done using the 
Multichannel Image Reconstruction, Image Analysis and Display package  
\citep[MIRIAD;][]{MIRIAD1995}\footnote{http://carma.astro.umd.edu/miriad/}.   
The bandpass and flux calibrators for each track are  
listed in Table \ref{tabObs}.  
%
After the data are reduced, 
the flux density of both primary and secondary phase calibrators is  
plotted as a function of $u$-$v$ distance in order to verify  
a flat trend, implying  
that decorrelation is not significant at long baselines. 

The largest uncertainty of interferometric data comes from 
flux or absolute amplitude calibration. 
Although independent of relative brightness and image quality for one track of data, 
it affects the analysis through the differences between tracks. 
The uncertainty can be larger than 10\%  
mostly due to the planetary model 
of the flux calibrator used in the data reduction process  
\citep[e.g.,][]{Moreno2002}.  
The large uncertainty cannot be avoided unless 
the planet modeling is improved. 
The flux uncertainty can  affect the analysis through (1) the
uncertainty between tracks at the same wavelength, and (2) the uncertainty
between tracks at different wavelengths.  The uncertainty of the
first kind can affect the deduced envelope and disk structure in
the modeling.  
To ensure its impact is minimized, 
we compare the flux of the common phase calibrator 1927+739 among tracks.  
At each wavelength, we verify that the flux value 
varies smoothly in time and is consistent with the flux reported in the  
standard CARMA/MIRIAD catalog. 
Also, data and model are compared at each visibility point, so  
the uncertainties are better preserved
compared to modeling using binned visibilities 
(see \S \ref{sec:mproc}). 
For the rest of the paper we consider the absolute flux uncertainty 
of the second kind for dual-wavelength analysis. 
A 10\% uncertainty for the absolute flux at each wavelength is adopted.  
It dominates errors in estimating spectral parameter, 
such as the dust opacity spectal index, 
as will be seen in    \S \ref{sec:result}.  
Other model parameters can be affected directly or indirectly through the
uncertainty of the spectral parameter. 
%
Other systematic uncertainty from instruments and calibrations may 
exist and propagate in the analysis as well, 
but are presumably less than 10\%.   

 
%
%

\begin{figure}
\includegraphics[angle=270,width=1.0\textwidth]{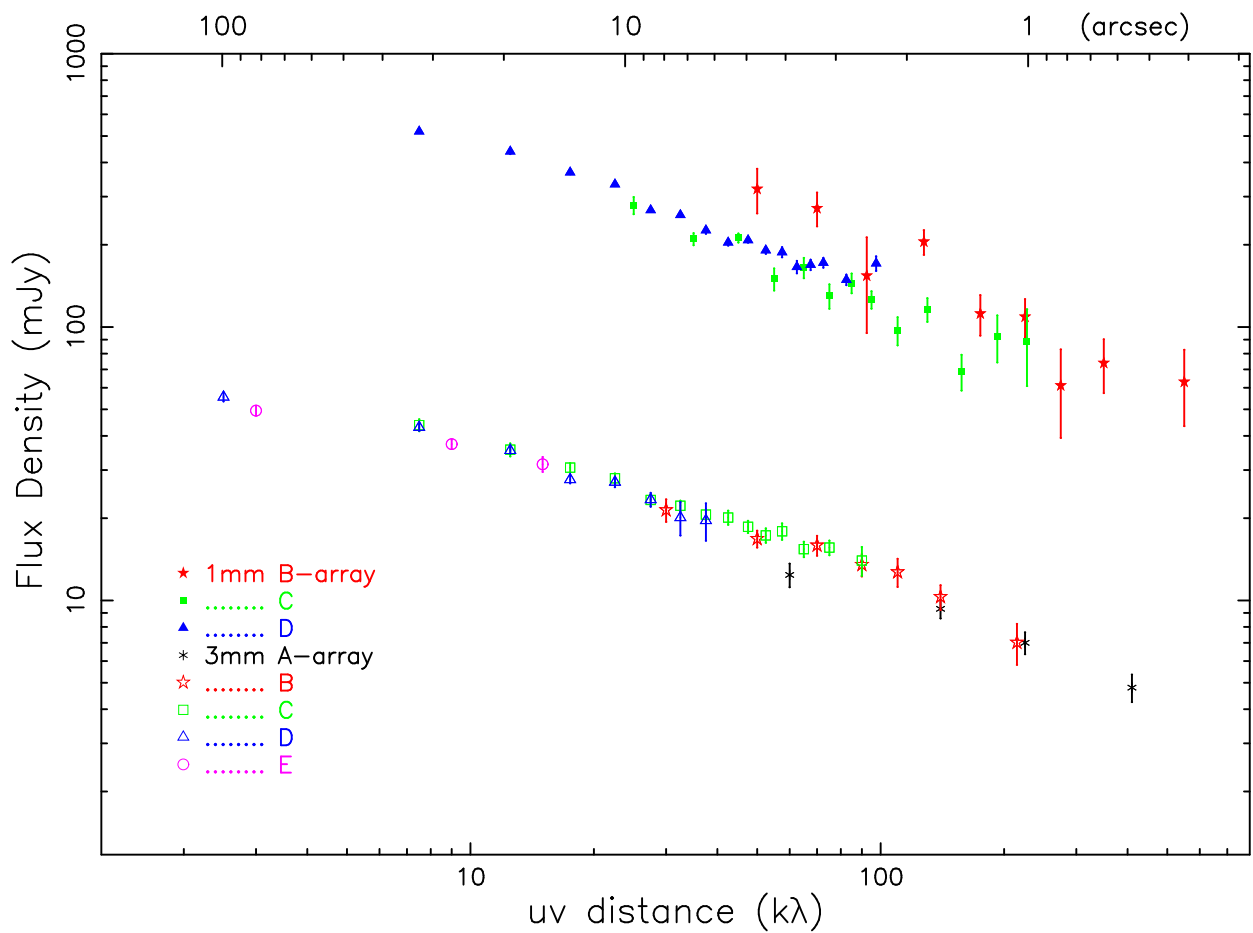}
\caption[Flux density of L1157-mm dust continuum at 
multiple array configurations. ]{
Flux density of L1157-mm dust continuum 
at 1 mm ({\it filled symbols}) and 3 mm ({\it open symbols}).  
The visibilities are 
vector averaged around the source center  
and binned into {\it u-v} annuli.  
Data collected in different array configurations are plotted separately. 
The error bars show only the statistical errors  within each annuli-bin, 
while the typical uncertainty carried by single data visibility 
is around 0.2-2.5 Jy. 
}
\label{figConfigs}
\end{figure}

\begin{figure}
\includegraphics[angle=270,width=1.0\textwidth]{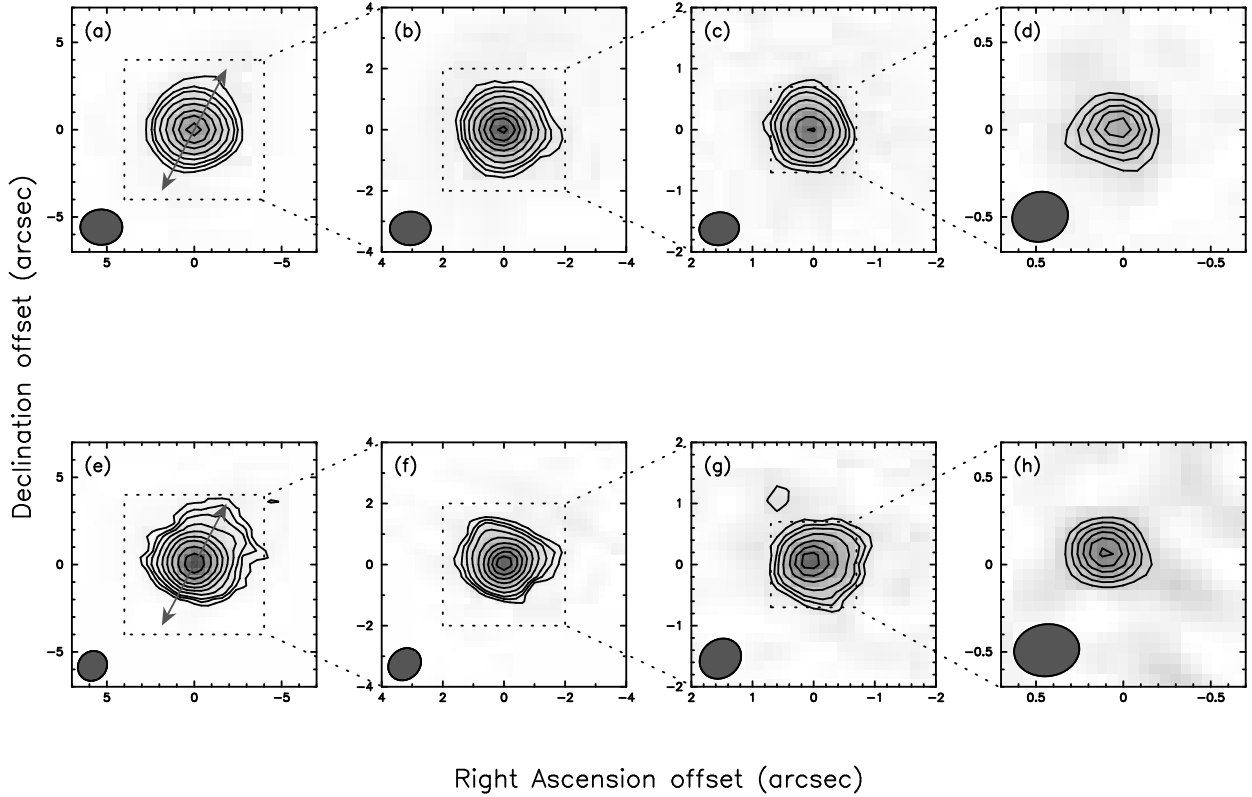}
\caption[CARMA 3 mm and 1 mm dust continuum images of L1157-mm. ]{ CARMA 3 mm ({\it upper}) and 1 mm ({\it lower}) 
dust continuum images of L1157-mm. 
The same multi-configuration data with different {\it u-v} imaging weightings 
are shown to emphasize structures on different size scales.
The contour levels, noise rms ($\sigma$), and beams are as follows: 
(a) [3,4,5,7,10,14,18,22]$\times \sigma$, $\sigma =$ 0.9 mJy beam$^{-1}$, 
2.40$\arcsec \times$2.03$\arcsec$ at a position angle of   90$^{\circ}$;
(b) [3,4,5,7,10,14,18,22,26]$\times \sigma$, $\sigma =$ 0.6 mJy beam$^{-1}$, 
1.34$\arcsec \times$1.10$\arcsec$ at a position angle of -88$^{\circ}$;
(c) [3,4,5,7,10,14,18]$\times \sigma$, $\sigma =$ 0.6 mJy beam$^{-1}$, 
0.65$\arcsec \times$0.54$\arcsec$ at a position angle of  -82$^{\circ}$;
(d) [3,4,5,6,7]$\times \sigma$, $\sigma =$ 0.9 mJy beam$^{-1}$, 
0.32$\arcsec \times$0.28$\arcsec$ at a position angle of  -73$^{\circ}$.
(e) [3,4,5,7,10,14,20,30,42]$\times \sigma$, $\sigma =$4.0 mJy beam$^{-1}$,  
1.77$\arcsec \times$1.61$\arcsec$ at a position angle of -38$^{\circ}$; 
(f) [3,4,5,7,10,14,18,22,26]$\times \sigma$, $\sigma =$5.5 mJy beam$^{-1}$, 
1.14$\arcsec \times$0.98$\arcsec$ at a position angle of -48$^{\circ}$;
(g) [3,4,5,7,10,13,16]$\times \sigma$, $\sigma =$ 7.0 mJy beam$^{-1}$, 
0.71$\arcsec \times$0.62$\arcsec$ at a position angle of -51$^{\circ}$;
(h) [3,4,5,6,7,8]$\times \sigma$, $\sigma =$ 12.0 mJy beam$^{-1}$, 
0.37$\arcsec \times$0.30$\arcsec$ at a position angle of -83$^{\circ}$.
 }
\label{figDustCont}
\end{figure}

The reduced data of the science target L1157-mm are shown 
in Figure \ref{figConfigs} by the annuli-averaged 
flux density with respect to $u$-$v$ distance. 
Continuum data from 
all spectral windows are combined. 
Figure \ref{figDustCont} presents the continuum maps of L1157-mm.  
Super-uniform weightings with different robustness parameters  
are used to obtain different synthesized beamsizes  
in order to emphasize envelope structures at different size scales.  
The continuum of L1157-mm shows spherical structures 
from 2000 AU scale (8\arcsec) down to 100 AU scale (0.4\arcsec).  
In panel (f), the envelope structure is slightly elongated 
perpendicular to the outflow direction, but the larger-scale extended 
envelope, detected by IRAM 30-m telescope \citep[][]{Gueth2003}   
and SMA (with lower resolution than our CARMA observations; 
Tobin et al. in preparation), 
is not seen in our CARMA dust continuum data. 
No apparent disk or flattened structure is seen at small scale either.  

\section{Aspects of Modeling a Class 0 YSO } \label{sec:modeling} 

To compare a YSO model with observations, 
we consider the physical conditions of the system, 
including the density and temperature structures 
(\S \ref{sec:envstruct} and \S \ref{sec:temp}) 
and dust grain properties (\S \ref{sec:dust}). 
The radiative transfer tool RADMC-3D,  
developed by C. P. Dullemond and co-authors \citep{RADMC2004}    
\footnote{http://www.ita.uni-heidelberg.de/$\sim$dullemond/software/radmc-3d/}, 
is used.  
Observational effects from the interferometer  
are taken into account 
and a Bayesian approach is taken for model fitting (\S \ref{sec:mproc}).  
In the following sections, we discuss the details 
of each facet in the modeling.  


\subsection{Envelope Structure }  \label{sec:envstruct} 

A simple Class 0 YSO model  
consisting of a spherical dusty envelope, a bipolar outflow, and 
possibly a circumstellar disk is considered. 
For the envelope structures, we examine  
a simple power-law density profile, representing self-similar 
collapse solutions,  
and a collapse with rotation (Terebey et al. 1984). 
An unresolved component is included to represent a compact disk 
structure.

To include a simple bipolar outflow cavity in the model, we 
remove material orientated with the observed
envelope geometry and outflow properties:
a position angle of 152\degr~for the outflow-axis cut, 
an inclination angle of 80\degr,  
and an opening angle 30\degr~for the outflow cavity 
are assumed \citep{Choi1999,Gueth1996,Gueth1997}. 
For simplicity, the inner and outer radii of the envelope are fixed to be 
12 AU and 10,000 AU, respectively.  
The inner envelope cavity is smaller than the highest 
observational resolution,  
and always within the central cell in the model images. 
A large outer radius is adopted so there is no ringing effect due to 
interferometric response on a sharp cutoff in the envelope. 
Additionally, as the density and temperature are much lower in the outer 
envelope, precise choice of the outer radius 
does not play an important role at these wavelengths.   


\subsection{Temperature Structure}\label{sec:temp}

While many theoretical models ignore the internal heating 
from the newborn protostar, it is critical to take into account 
the protostellar contribution to agree with observational luminosity 
\citep{Adams1985}.  
The heating and cooling of dust grains, 
dominated by the central illumination and dust grain properties, 
should be balanced to obtain an equilibrium temperature.   
To simulate millimeter-wave observations of protostars surrounded by dusty environments, 
such a realistic temperature distribution needs to be either assumed or calculated.  

The temperature structure 
can be approximated assuming simple conditions of the dusty envelope. 
Assuming 
a centrally illuminated spherical envelope in which the density has a power-law 
dependence on radius,   
\begin{equation}
\rho(r)   = \rho_0  \left(\frac{r}{r_0} \right)^{-p},   
\label{eqPowlDen}
\end{equation}
where $\rho_0$ is the density at an arbitrary radius $r_0$ and 
$p$ is the density power-law index, 
and assuming a pure power-law dust opacity with a spectral index $\beta$,  
the temperature structure in the optically thin outer envelope 
can be approximated by  
\begin{equation}
T(r)    = T_0  \left(\frac{r}{r_0} \right)^{-\frac{2}{\beta+4}}  
\label{eqTemp}
\end{equation}
\citep{WC1986,Adams1991}. 

However, if the assumptions of power-law dust opacity 
and spherical power-law density  do not hold, 
the approximation in Eq.~(\ref{eqTemp}) can be inadequate  
even in the optically thin reion. 
For example, 
the dust opacity is not a pure power-law 
at short wavelengths, and 
the density structure can also be more complicated than the power-law profile.  
Furthermore, Eq.~(\ref{eqTemp}) is  
only valid in the optically thin region and relies on $T_0$ at $r_0$.  
With a fixed central heating source, $T_0$ at $r_0$ is characterized 
by the optically thick-thin transition zone, 
and is difficult to estimate 
without a good understanding of the optically thick inner region. 
Given the difficulty to approximate the temperature structure 
with a variety of envelope models and ranges of model parameters, 
we calculate a self-consistent temperature distribution  
for each set of parameters 
using the Monte Carlo radiative transfer code RADMC-3D \citep{RADMC2004}. 
A luminosity of 8.4 L$_\odot$ is adopted 
\citep{Froebrich2005} 
as a fixed input in the radiative transfer calculation.  
This bolometric luminosity can be underestimated 
due to insufficient sampling of the spectral energy distribution, 
but can be overestimated due to a larger assumed distance.  
Furthermore, the intrinsic luminosity  
can be larger than the measured bolometric luminosity 
due to the source's edge-on orientation   
\citep[e.g., see more discussions in][]{Froebrich2005,Whitney2003a}. 
Despite the uncertainty in luminosity, 
a self-consistent temperature structure 
is the best compromise for now.  

\subsection{Dust Grain Properties}\label{sec:dust} 

Dust grain properties such as chemical composition, geometry, 
alignment, degree of ionization, and size distribution 
play important roles in star-forming processes 
from thermodynamics and grain surface chemistry  
to timescales of magnetic field effects.   
For dust grains in the diffuse interstellar medium, 
the classic model constructed by \citet*[][hereafter MRN]{MRN1977}
with optical constants calculated by \citet{DraineLee1984}  
can reproduce 
the interstellar extinction and polarization observations  
from infrared to ultraviolet wavelengths.   
However, for dust grains in dense cores and star forming regions, 
collisions and interactions between grain particles 
become more important. 
\citet{OH1994} has considered the dust coagulation process 
in dense protostellar cores 
and found that the opacity can be enhanced by a factor of a few  
as grains aggregate.  
The authors started with the MRN grains covered with different amounts of  
ice mantles, and investigated the optical constants after 
$10^5$ yrs of coagulation in gas densities ranging from $10^5$ to 
$10^8$ cm$^{-3}$. 
In our modeling, we adopt %
the dust opacity, or the mass absorption coefficient 
$\kappa$ defined as the cross section per unit mass,   
from column 5 of Table 1 in \citet{OH1994}, 
the so-called OH5 grain which is covered by a 
thin layer of ice mantle and coagulated at $10^6$ cm$^{-3}$.   
Besides being widely used, 
the OH5 model shows agreements with, and in some cases favored by, 
multi-wavelength observations of star-forming regions  
\citep[e.g.,][]{vanderTak1999,Evans2001,Shirley2005,Shirley2011}. 

At far-infrared and millimeter wavelengths, 
$\kappa$ can be approximated as a power law with respect to frequency  
\begin{equation}
\kappa = \kappa_0  \left(\frac{\nu}{\nu_0} \right)^\beta .  
\label{eqKappa} 
\end{equation}
This sub-millimeter dust opacity spectral index $\beta$, 
which can only be studied with multi-wavelength observations,
varies with environment and is related to grain properties 
mentioned previously. 
$\beta$ is $\gtrsim$ 1.7 in the diffuse interstellar medium and starless cores 
\citep[e.g.,][]{DraineLee1984,Schnee2010}, 
but significantly lower in protoplanetary disks \citep[$\beta~\lesssim~$1, e.g.,][]{Beckwith1991,Natta2007PPV,Ricci2010a}.
One explanation for lower $\beta$ is a larger grain size 
and more discussions will be in \S 5.4. 
In order to better understand the dust property of L1157-mm,  
we include $\beta$ as a model parameter.  
Based on the OH5 model, we modify the opacity curve 
with a power-law of %
index $\beta$ as in Eq.~(\ref{eqKappa})    
at wavelengths longer than an arbitrary choice of 700 $\mu$m. 
The dust opacity used in the analysis 
can substantially affect %
the deduced spectral energy distribution 
as well as temperature structure of YSOs. 
With the radiative transfer tool, we find that 
the temperature structure is mostly determined by 
the dust opacity at short wavelengths.   
$\beta$, which characterizes the dust property at long wavelengths, 
plays a less important role  for the temperature structure;
instead, $\beta$ affects the observed flux directly through dust opacity 
\citep{Chandler1998}. 

\begin{figure}
\includegraphics[angle=0,width=1.0\textwidth]{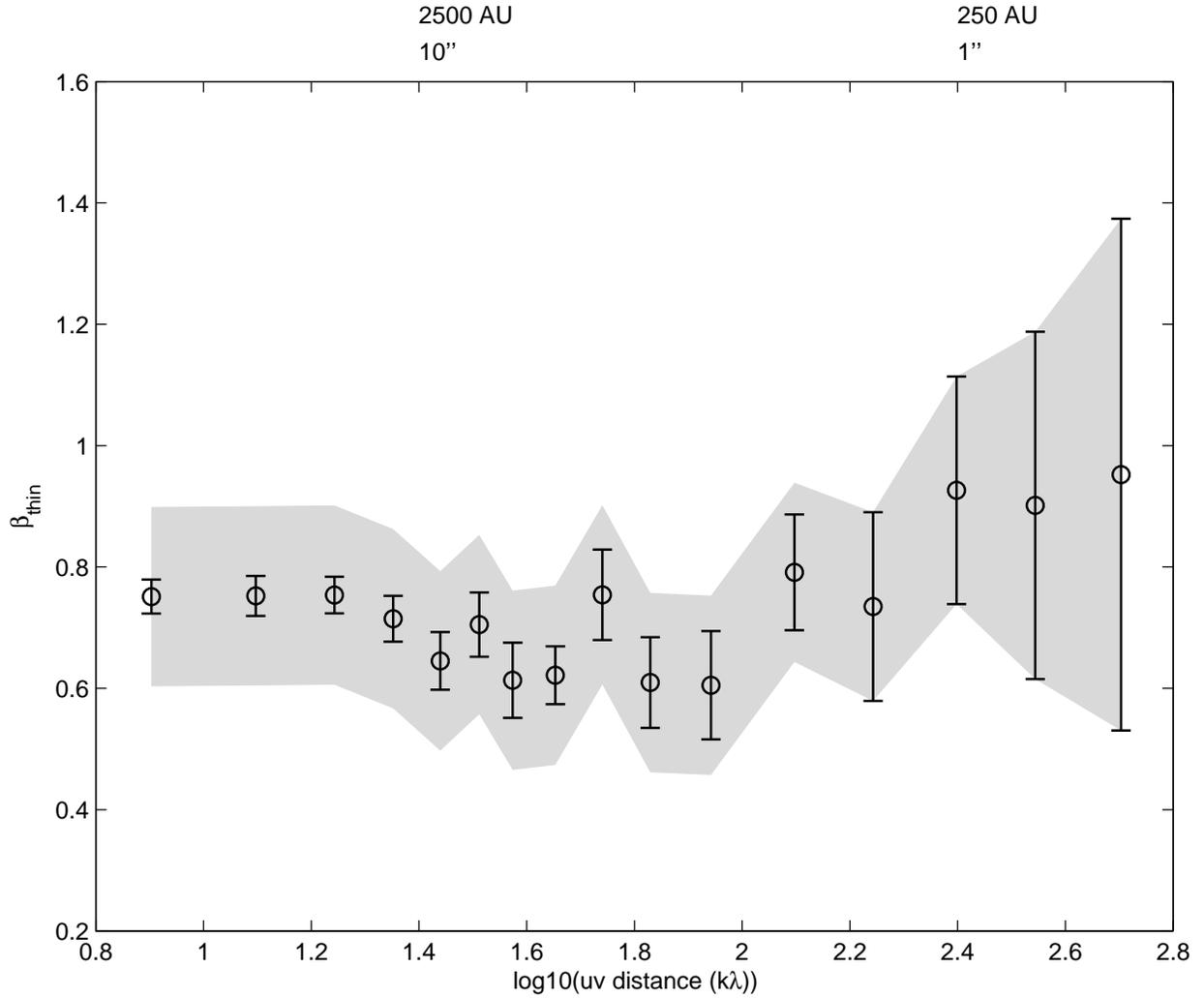}
\caption[Approximate dust opacity spectral index of L1157-mm 
as a function of \emph{u-v} distance 
assuming optically thin condition. ]{
Approximate dust opacity spectral index $\beta_{thin}$ of L1157-mm 
as a function of \emph{u-v} distance 
assuming the optically thin condition. 
The error bars show only the statistical errors without the absolute 
flux uncertainty, while  
the shade shows the errors including the absolute flux uncertainty.  
 }
\label{figBeta}
\end{figure}

With the optically thin assumption, %
$\beta$ can be estimated 
using flux ratio between two wavelengths in the Rayleigh-Jeans regime; 
here we call the approximate dust opacity spectral index $\beta_{thin}$. 
In this limit, the flux density  
$F_\nu \propto \kappa_\nu B_\nu \propto \nu^{\beta_{thin}+2}$;  
therefore, 
\begin{equation} 
\beta_{thin} = \frac{\ln F_1 - \ln F_2}{\ln \nu_1 - \ln \nu_2} - 2   
\label{eqBeta} 
\end{equation} 
\citep[e.g.,][]{Kwon2009}. 
Figure \ref{figBeta} shows  $\beta_{thin}$  of L1157-mm  
using the annuli-averaged visibility at each \emph{u-v} distance bin 
of our dual-wavelength data.  The uncertainty of 
the $\beta_{thin}$ estimation is discussed in Appendix A.   
$\beta_{thin}$ is a good approximation in the optically thin region, but 
a correction term is needed in the optically thick region 
\citep[e.g.,][]{Rodmann2006,Lommen2007}. 
To avoid the need of the correction term, 
full optical depth effect is considered  
in our radiative transfer modeling. 
Nonetheless, $\beta_{thin}$ 
provides a quick and rough estimate for the $\beta$ value across the envelope   
and reveals possible radial dependence.  
As seen in Figure \ref{figBeta}, no strong radial dependence is suggested for $\beta$ at L1157-mm. 
Therefore, we assume uniform grain properties in the envelope model for simplicity. 
In other words, $\kappa$ is only a function of frequency and 
independent of radius in our envelope model.  

\subsection{Free-free Contamination }
We ignore the contribution of free-free emission in this study. 
Free-free emission from ionized winds or jets can contribute 
partial flux at millimeter wavelengths 
\citep[$\sim$20\% at 7 mm,][]{Rodmann2006}  
and affect model parameter estimates especially for $\beta$ and disk component.  
However, it plays a minimal role for our data of L1157-mm at 1 mm and 3 mm. 
By extrapolating fluxes at 8.5 GHz and 4.86 GHz \citep{Meehan1998} 
to our observed frequency, we estimated the free-free emission  
to be around 0.53 mJy at 3 mm and 0.39 mJy at 1 mm for L1157-mm. 
The free-free correction is negligible in the analysis.    


\subsection{Model Fitting}\label{sec:mproc}  

Given a set of model parameters, we 
estimate the sky brightness distribution for the dust continuum 
with radiative transfer calculations.  
The density and self-consistent temperature distributions in 
three dimensions are considered along with the model dust property.   
Essentially, 
for each pixel on the plane of sky, 
the flux is calculated by integrating 
the dust emission along the line of sight \citep[e.g.,][]{Adams1991}.   
Assuming no background brightness, 
the specific intensity can be expressed as 
\begin{equation}   
 I_\nu = \int_{los} B_\nu(T) \, e^{-\tau_\nu} \, \textrm{d}\tau_\nu 
       = \int_{los} B_\nu(T(\vec{r})) \, e^{-\tau_\nu(\vec{r})} \,  
         \rho(\vec{r}) \, \kappa_\nu \, \textrm{d}\vec{r},
\end{equation}     
where $B_\nu(T)$ is the Planck function at dust temperature $T$, 
$\rho$ is the envelope density,  
$\vec{r}$ denotes the position,  
and $\tau_\nu$ is the optical depth from the position $\vec{r}$ along the 
line of sight ($los$) to the observer  
\begin{equation}   
\tau_\nu(\vec{r}) = \kappa_\nu \int_{los} \rho(\vec{r}) d\vec{r} 
                  = \kappa_\nu \int_l^\infty \rho(\vec{r}) dl'  .  
\end{equation}     
$T$, $\rho$, and $\tau_\nu$ are all dependent of $\vec{r}$.   

With the model sky brightness, we simulate interferometric observations 
and generate model visibilities.  
The sky image is convolved with the primary beam patterns of the antennas  
and then Fourier transformed into visibilities with the observational
{\it u-v} sampling.  
In the case of the 15-element CARMA, the 6.1-meter dishes and 10.4-meter 
dishes give 3 types of baselines. 
Therefore we construct separate primary-beam-corrected images for each
kind of baseline, and sample the images with corresponding {\it u-v}
spacing for each data visibility from real observations.
In addition, images at two wavelengths are constructed individually 
based on the same model.  

Model visibilities are compared with observational data 
at each {\it u-v} sample and wavelength. 
The analysis is done in the visibility domain 
so as to avoid the complexity brought by the CLEAN algorithm, 
{\it u-v} sampling, and imaging process. 
Some information is lost in the image domain through the imaging process, 
since structures in images can be sensitive to beamsize or weighting. 
In other words, emission at different size scales can either be 
emphasized or suppressed, 
causing biases in the model-data comparison; 
therefore, we perform the analysis in the visibility domain. 
Furthermore, visibilities are compared data point by data point; 
no binning nor averaging are done 
\citep[e.g., ][]{Isella2009}. 







Assuming the noise from observations is  normally distributed  
or Gaussian noise, 
the goodness of a model-fit can be characterized by 
the standard chi-square statistics.  Real and imaginary parts of each 
visibility point are considered individually, 
as in   
\begin{equation}  \label{eq:chi2}
 \chi^2 = \displaystyle{\sum_i} 
 \frac{ (Re(V_{model,i})-Re(V_{data,i}))^2 + 
        (Im(V_{model,i})-Im(V_{data,i}))^2  }{\sigma_i^2}    
\end{equation}
where $i$ stands for each visibility point at its unique $u$-$v$. 
The noise of each visibility $\sigma$ is the square root of  
data variance (outputted by MIRIAD task \textsf{uvinfo})
multiplied by a scaling factor to account for imperfect weather conditions. 
The noise level before the scaling follows 
\begin{equation}
 \sigma_o  = \frac{2k_b T_{sys} }{\eta_a \eta_c A  \sqrt{N(N-1)\Delta \nu t_{int}} } , 
\end{equation}
%
where $k_b$ is the Boltzmann constant, 
$T_{sys}$ is the system temperature, 
$\eta_a$ is the aperture efficiency, 
$\eta_c$ is the correlator efficiency, 
$A$ is the antenna collecting area, 
$N$ is the number of antennas,
$\Delta \nu$ is the bandwidth, 
and $t_{int}$ is the on-source integration time.   

The scaling factor is used to correct $\sigma$ for the phase 
decorrelation and is 1 if no scaling is done.  
There are multiple ways to scale the noise, and the scaling 
factor should be somewhat dependent on the baseline length. 
In this work we determine the scaling factor by the phase scatter  
in each array configuration at each wavelength.  
Nonetheless, the factor is always larger than 1; 
in other words, 
we only adjust $\sigma$ to make data less constraining.


We take the Bayesian approach to compare data and  model  
\citep[e.g.,][]{Ford2005,Spergel2007}.  
Given a specific model, 
a global minimum of $\chi^2$ is searched and verified to be 
a good fit with a chi-square hypothesis test. 
Then, 
the Markov chain Monte Carlo (MCMC) method 
is utilized to calculate the posterior probability distributions;  
in particular, 
the posterior-weighted value and the uncertainty 
are estimated for each model parameter.  
Details of our fitting technique are discussed in Appendix B. 
Note that the deduced parameters 
are valid only within the 
framework of model assumptions. 
Evaluating the goodness of a model  and  
comparisons between models will be discussed in \S \ref{sec:modelSelection}.  


\section{Results}  \label{sec:result} 

In this section, the modeling results 
based on the details described 
in \S \ref{sec:modeling} are presented.  
Three   models are considered and 
shown individually. %

\subsection{Spherical Power-law Envelope Model}\label{sec:m1} 

We first consider a spherical envelope 
with a power-law density profile and self-consistent 
temperature structure.  In this simplest model, 
three model parameters are included: 
(a) the dust opacity spectral index $\beta$ as in Eq.~(\ref{eqKappa}),   
(b) the dust density $\rho_0$ at 100 AU, 
which scales with the total envelope mass,  
and (c) the density power-law index $p$ as in Eq.~(\ref{eqPowlDen}). 
All other model properties are fixed 
as described in \S \ref{sec:modeling}.

\begin{figure}
\includegraphics[angle=0,width=1.0\textwidth]{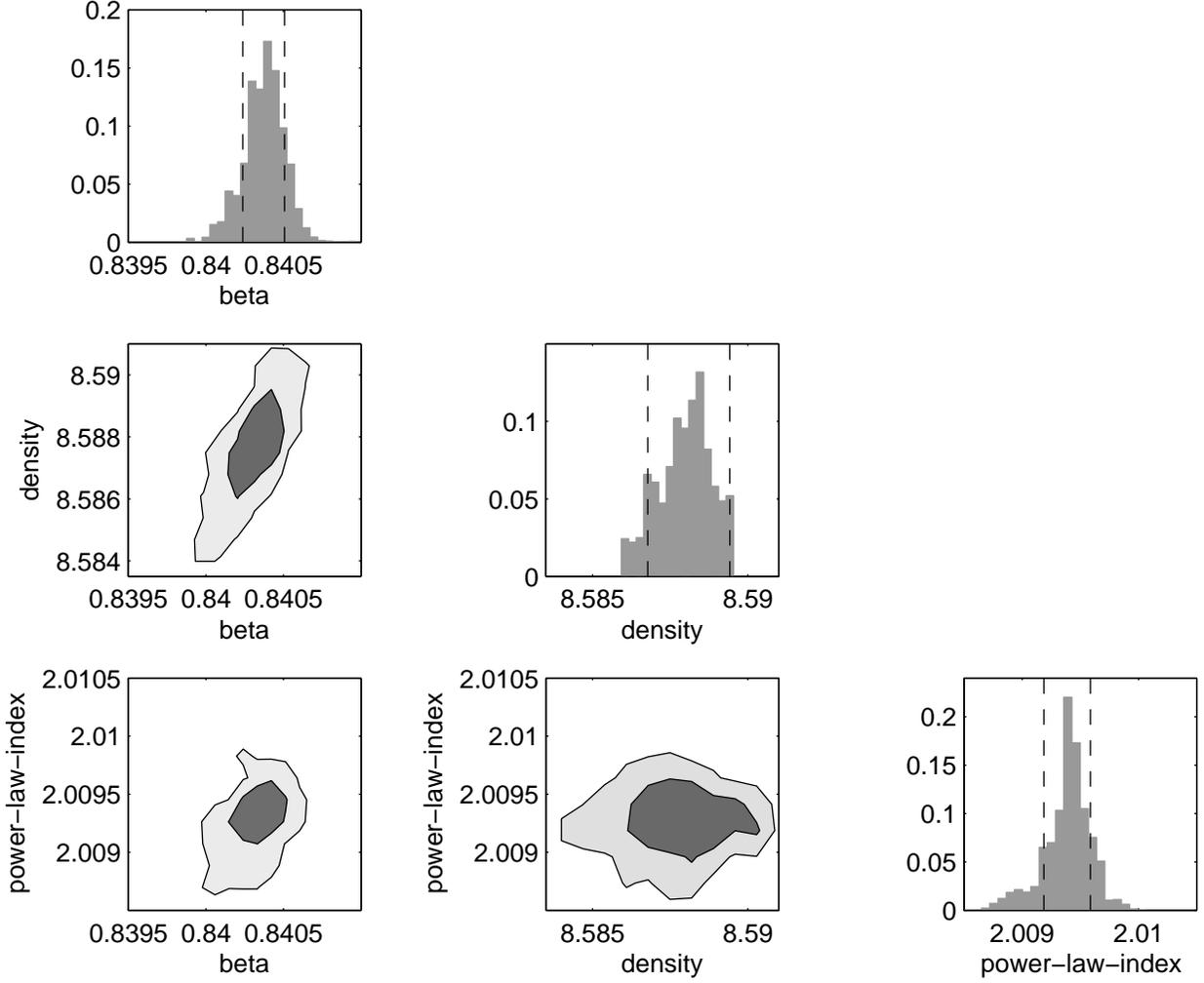}  
\caption[Marginalized posterior probability distributions of 
the power-law model without absolute flux uncertainty. ]{ 
Marginalized posterior probability distributions for the parameters
of the power-law envelope model
(dust opacity spectral index $\beta$, 
density $\rho_0$ at 100 AU in the unit of 10$^{-18}$~g~cm$^{-3}$, 
and envelope density power-law index $p$).  
Only the statistical errors, but not the absolute flux uncertainty,  
are considered for the data. 
In the
histograms, the dashed vertical lines enclose 68\% or 1 $\sigma$
confidence interval, with the expectation values and $\sigma$ listed in
Table \ref{tabM1}.  The dark and light areas in the 2-D contour plots
are the 68\% and 95\% confidence regions.
}
\label{figPDFm1}
\end{figure}

We begin with only considering the statistical noise of data visibility 
and ignoring the uncertainty of absolute flux.   
This is to characterize  the case without absolute flux uncertainty, 
as well as demonstrate the effect of absolute flux uncertainty.  
Using the MCMC results, 
the expectation values and uncertainties of all parameters 
are estimated (Table \ref{tabM1}),  
and the marginalized posterior probability distributions
in 1-D and 2-D parameter space are shown in Figure \ref{figPDFm1}.  
We choose to list the radius of the 68\% confidence interval  
in Table \ref{tabM1} as it represents 1 $\sigma$.  
The 68\% and 95\% confidence regions 
are shown by 2-D contours to reveal any correlation between parameters.  
For example, as shown by the marginalized contours in the 
$\beta$-$\rho_0$ plane in Figure \ref{figPDFm1} (middle row, left column),  
a correlation between 
the dust opacity spectral index $\beta$ and 
the density scaling $\rho_0$  
is implied. 
This is because a larger $\beta$ means a smaller dust opacity $\kappa$ 
in the model, resulting in larger deduced mass and thus larger density scaling. 

\begin{deluxetable}{lccc} 
\tabletypesize{\small}
\tablecolumns{4} 
\tablewidth{0pc} 
\tablecaption{Power-law Envelope Model} 
\tablehead{ 
\colhead{Parameter}  &  
\colhead{Mean }  &  
\multicolumn{2}{c}{Radius of the 68\% confidence interval} 
\\
\colhead{}  &  
\colhead{}  &  
\colhead{(statistical noise only)}  &  
\colhead{(with flux uncertainty)}   
\\
\colhead{(1)}  & 
\colhead{(2)}  & 
\colhead{(3)}  & 
\colhead{(4)}    
}
\startdata 
dust opacity spectral index $\beta$ \dotfill & 
0.84037 & 0.00014 & 0.11 
\\
density at 100 AU (in 10$^{-18}$~g~cm$^{-3}$)  \dotfill & 
 8.5881\tablenotemark{a} & 0.0013  & 1.2   
\\
density power-law index $p$  \dotfill & 
2.00939 & 0.00020 & 0.03
\\
\enddata 
\tablenotetext{a}{
corresponding to a total envelope mass of 1.8120 M$_\odot$ 
}
\label{tabM1}
\end{deluxetable}

The parameters are determined with a high precision 
within the framework of model assumptions.  
The narrow uncertainties can be understood since 
approximately five million independent data visibilities 
are used to fit only three model parameters.  
Strong assumptions are imposed in the model. 
Similar results have been obtained in other studies as well. 
For example, \citet{Kwon2011} have taken the Bayesian approach   
to estimate model parameters and their errors in the applications 
of T-Tauri disks, and small uncertainties are 
obtained when only the statistical errors in the data are considered. 

However, the absolute amplitude uncertainty, originated by the absolute
flux calibration in the data reduction process, brings more uncertainties
to the model parameter estimation. 
As we have verified that the calibrator flux is consistent among all
observational tracks at multiple array configurations (\S \ref{sec:obs}), 
the absolute flux errors effectively cause an uncertain scaling to the amplitude of all data at  
each wavelength.  The absolute flux uncertainty can play a
dominating role in estimating frequency-dependent parameters, but cause
minimal effects at the relative spatial structures
probed at one single wavelength. 

Marginalization in the framework of Bayesian statistics 
allows us to quantitatively   
take the absolute flux uncertainty into consideration. 
We introduce two additional nuisance parameters, $S_{1mm}$ and $S_{3mm}$, 
to scale the absolute amplitude of all data 
at 1 mm and 3 mm, respectively. 
$S_{1mm}$ = 1 and $S_{3mm}$ = 1 means no scaling is done, 
as in the presented dataset.  
Since the main uncertainty of flux calibration 
results from the choice of planetary models and 
no model is preferred, 
a flat probability distribution for both $S_{1mm}$ and $S_{3mm}$,  
ranging from 0.9 to 1.1, is assumed.  
The range of the scaling factors is chosen to be consistent with the 
commonly quoted 10\% errors for the absolute flux calibration. 
Our approach is similar to the method of \citet{Lay1995}. 

\begin{figure}
\includegraphics[angle=0,width=1.0\textwidth]{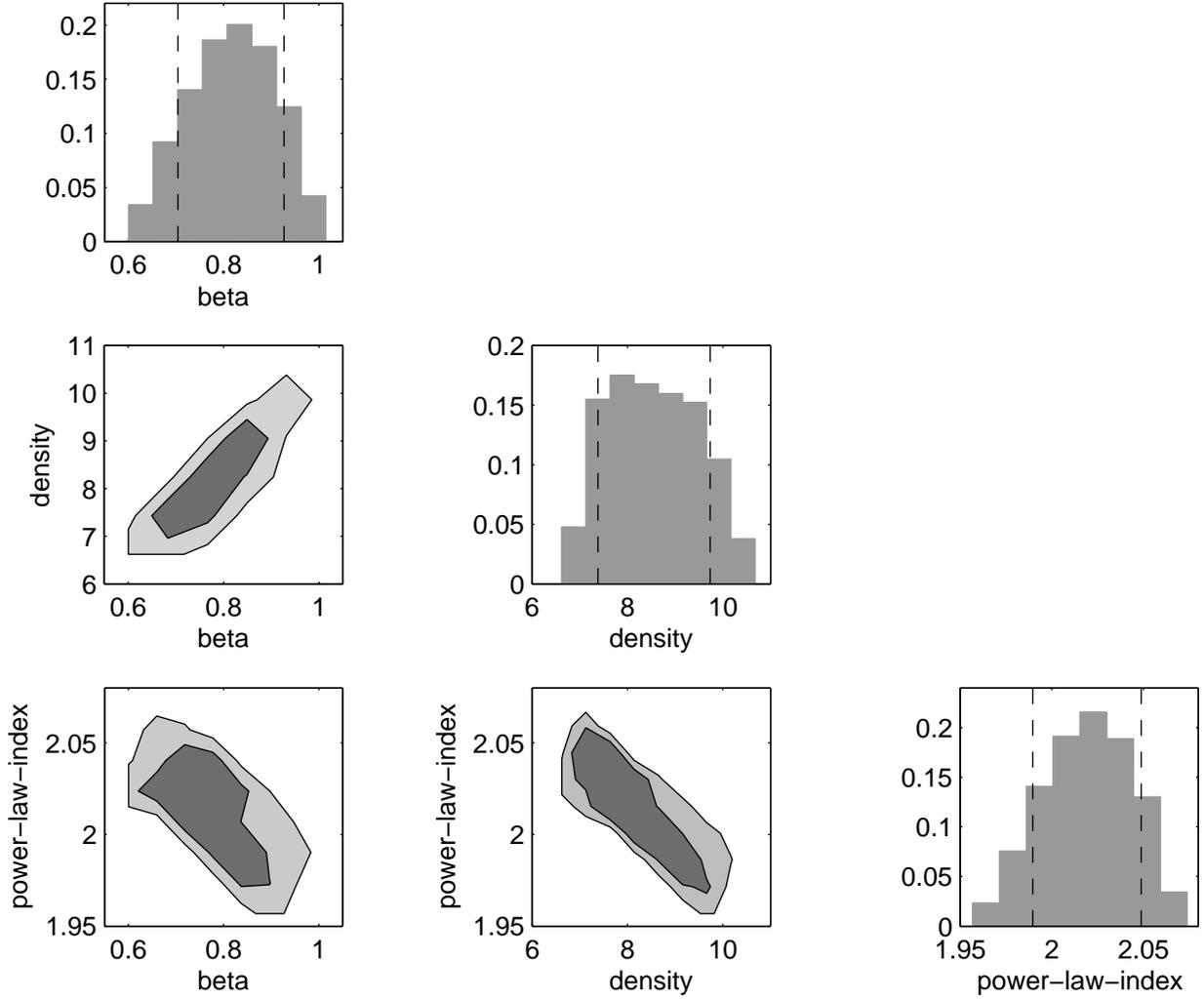}  
\caption[Marginalized posterior probability distributions of 
the power-law model with absolute flux uncertainty. ]{ 
Same as Figure  \ref{figPDFm1} but with the consideration of 
the absolute flux uncertainty.  
$\sigma$ for the model parameters 
are listed in Table \ref{tabM1} column 4.  
}
\label{figPDFm1_fs}
\end{figure}

Figure \ref{figPDFm1_fs} shows the marginalized posterior probability 
distributions of model parameters with consideration of the absolute 
flux uncertainty. The uncertainties of parameters are listed in 
Table \ref{tabM1} column 4.  
Inclusion of absolute flux uncertainty increases the 
parameter errors by a factor of 2-3 orders of magnitude, and 
it is critical for parameter   
estimation as it makes data much less constraining. 
In particular, the dust spectral index $\beta$ is mostly 
determined by the flux ratio between 1 mm and 3 mm, hence it becomes   
much more uncertain due to the uncertainty of absolute amplitude.  

\begin{figure}
\includegraphics[angle=270,width=0.95\textwidth]{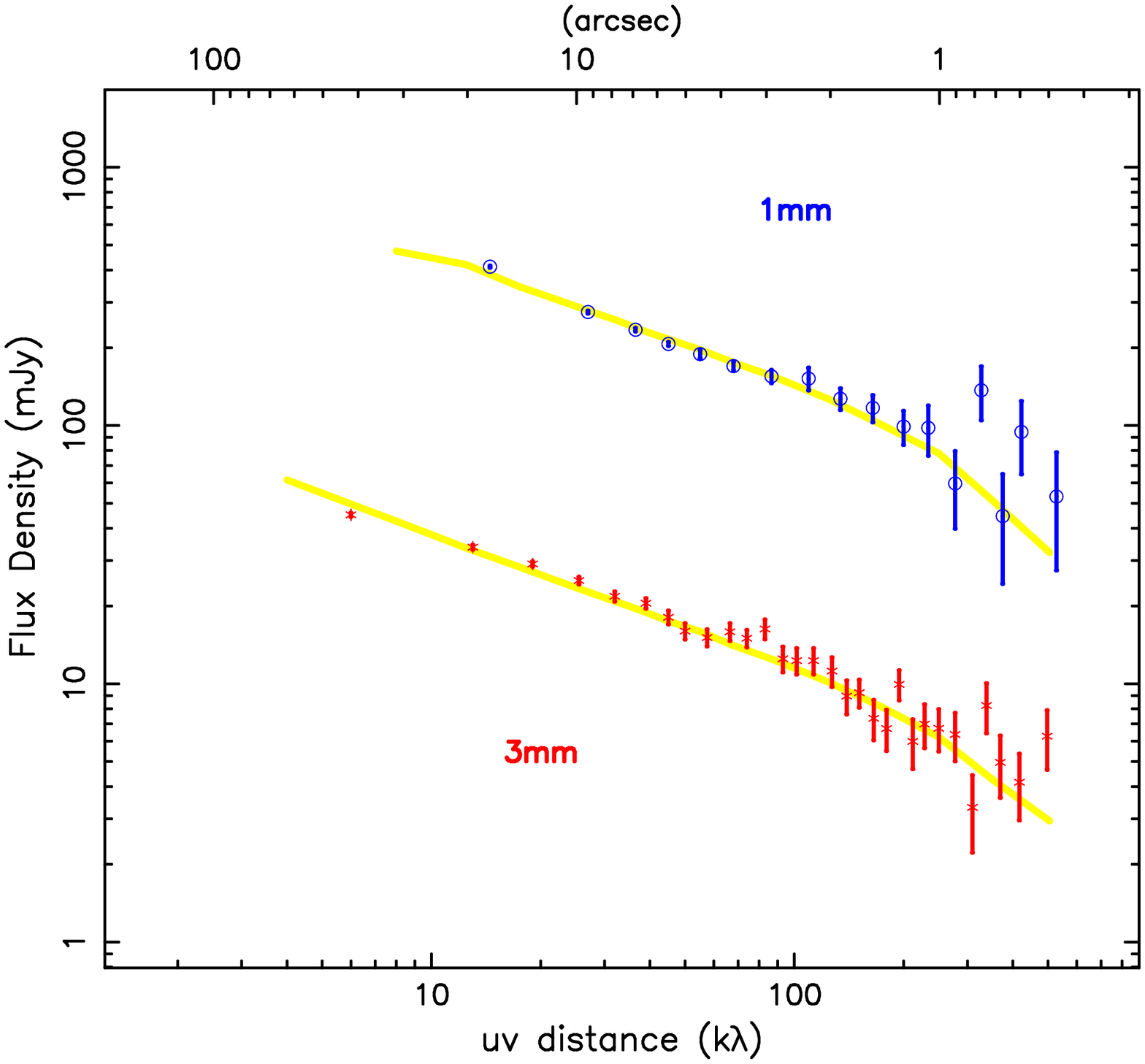}
\caption[Visibility comparison of the power-law model. ]{ 
Flux density of the observational data 
(circles for 1 mm data and asterisks for 3 mm data)  
and the model fit (solid lines) with the marginalized parameters 
for the power-law envelope model.     
While modeling is done with non-averaged visibilities, annuli-averaged 
visibilities are shown as a function of $u$-$v$ distance. 
Error bars are statistical errors in the annuli-bins only 
and different from the visibility uncertainty used in the modeling. 
}
\label{figBinM1}
\end{figure}

For additional visualization, 
Figure \ref{figBinM1} shows 
the observational data visibility of L1157-mm and the model calculated with the
marginalized parameters 
as an {\it a posteriori} comparison. 
Because there are about five million data visibilities 
and each visibility contains low signal-to-noise, 
plotting them all does not show information.  
Therefore, visibilities are averaged vectorially and  
binned in $u$-$v$ annuli around the source center  
using the MIRIAD task {\textsf{uvamp}}.  
The error bars in Figure \ref{figBinM1} are statistical errors in the bins.  %
Note that visibility data are not averaged in the modeling process, 
and the binned visibility is just one data representation. 
The same dataset can have multiple representations depending on 
how they are binned, and the statistical errors in the bins 
may not reflect the whole uncertainty. %

\begin{figure}
\includegraphics[angle=270,width=1.0\textwidth]{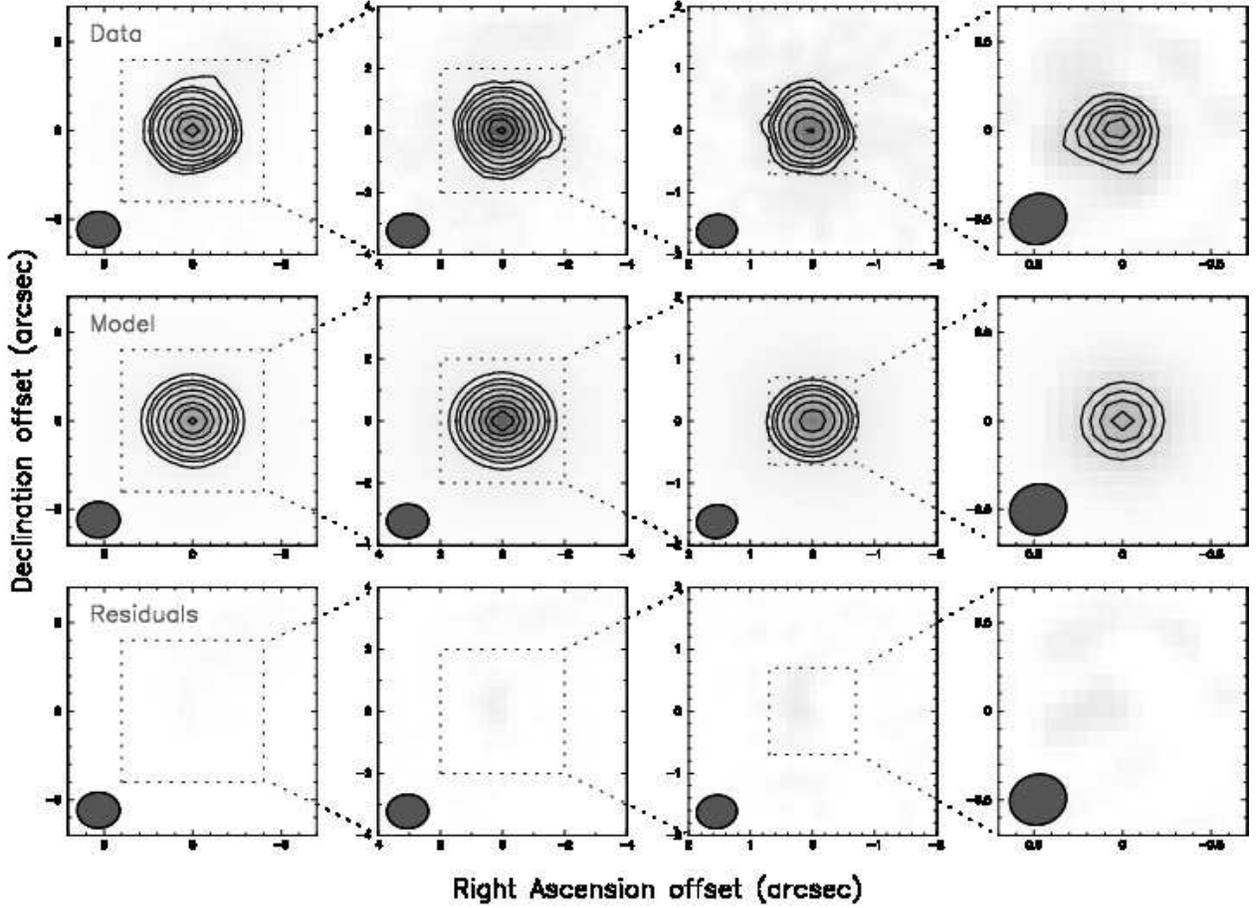}
\caption[Image comparison between data, model, and residuals 
at 3 mm. ]{ 
Comparison between 3 mm dust continuum data ({\it upper row}, 
as shown in Figure \ref{figDustCont}),  
model ({\it middle row}), and residuals ({\it lower row}) 
of L1157-mm in the image space. 
The power-law envelope model is used.
Images in each column share the same {\it u-v} imaging weighting and 
contour levels. 
The contour levels, noise rms ($\sigma$), and beams are: 
{\it column 1}:  [-3,3,4,5,7,10,14,18,22]$\times \sigma$, $\sigma =$ 0.9 mJy beam$^{-1}$, 
2.40$\arcsec \times$2.03$\arcsec$ at a position angle of   90$^{\circ}$;
{\it column 2}:  [-3,3,4,5,7,10,14,18,22,26]$\times \sigma$, $\sigma =$ 0.6 mJy beam$^{-1}$, 
1.34$\arcsec \times$1.10$\arcsec$ at a position angle of -88$^{\circ}$;
{\it column 3}:  [-3,3,4,5,7,10,14,18]$\times \sigma$, $\sigma =$ 0.6 mJy beam$^{-1}$, 
0.65$\arcsec \times$0.54$\arcsec$ at a position angle of  -82$^{\circ}$;
{\it column 4}:  [-3,3,4,5,6,7]$\times \sigma$, $\sigma =$ 0.9 mJy beam$^{-1}$, 
0.32$\arcsec \times$0.28$\arcsec$ at a position angle of  -73$^{\circ}$.
 }
\label{figImg3mmM1}
\end{figure}

\begin{figure}
\includegraphics[angle=270,width=1.0\textwidth]{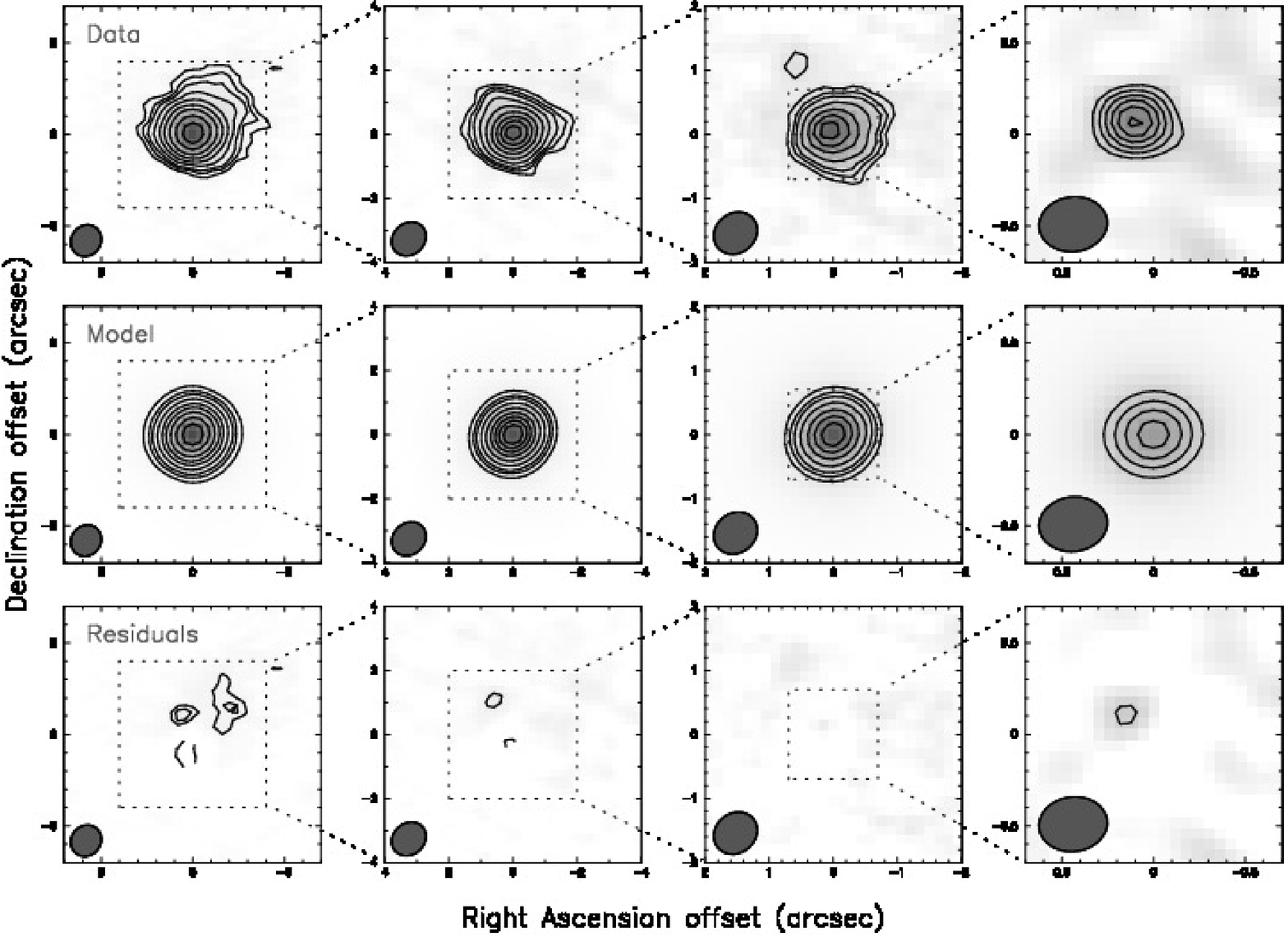}
\caption[Image comparison between data, model, and residuals
at 1 mm. ]{ 
Comparison between 1 mm dust continuum data ({\it upper row}, 
as shown in Figure \ref{figDustCont}),  
model ({\it middle row}), and residuals ({\it lower row}) 
of L1157-mm in the image space. 
Images in each column share the same {\it u-v} imaging weighting and 
contour levels. 
The contour levels, noise rms ($\sigma$), and beams are: 
{\it column 1}: [-3,3,4,5,7,10,14,20,30,42]$\times \sigma$, $\sigma =$4.0 mJy beam$^{-1}$,  
1.77$\arcsec \times$1.61$\arcsec$ at a position angle of -38$^{\circ}$; 
{\it column 2}: [-3,3,4,5,7,10,14,18,22,26]$\times \sigma$, $\sigma =$5.5 mJy beam$^{-1}$, 
1.14$\arcsec \times$0.98$\arcsec$ at a position angle of -48$^{\circ}$;
{\it column 3}: [-3,3,4,5,7,10,13,16]$\times \sigma$, $\sigma =$ 7.0 mJy beam$^{-1}$, 
0.71$\arcsec \times$0.62$\arcsec$ at a position angle of -51$^{\circ}$;
{\it column 4}: [-3,3,4,5,6,7,8]$\times \sigma$, $\sigma =$ 12.0 mJy beam$^{-1}$, 
0.37$\arcsec \times$0.30$\arcsec$ at a position angle of -83$^{\circ}$.
The 3 $\sigma$ residuals at the smallest-scale image ({\it lower-right})
is likely due to the differences  
of the emission peak position measured using 1 mm and 3 mm data, 
while the modeling is done around the peak position measured using 3 mm data. 
If we shift the model with this offset at 1 mm, no residuals 
is left at 3 $\sigma$ level   
in the small-scale image. 
 }
\label{figImg1mmM1}
\end{figure}

Figure \ref{figImg3mmM1} and Figure \ref{figImg1mmM1} 
continue the {\it a posteriori} check and   
compare the model with the data in the image domain 
for the 3 mm and 1 mm dust continuum, respectively.  
We image the model visibilities 
in the same way the data visibilities are imaged 
as shown in Figure \ref{figDustCont}, that is,   
the same sets of {\it u-v} imaging weightings are used for showing 
structures at four size scales at each wavelength.
Residuals in the visibility domain are also imaged and shown in 
Figure \ref{figImg3mmM1} and Figure \ref{figImg1mmM1} 
to demonstrate the fitting error in the image space.   
The subtraction of model from data 
leaves no residuals greater than 3 $\sigma$ level at the 3 mm images, 
confirming that a good fit is obtained.   
In the large-scale image of 1 mm continuum, 
the residuals extend towards the north-west of the protostar,  
which aligns with the outflow direction and 
is likely due to the asymmetric structure from the outflow.   
In the small-scale image of 1 mm continuum, 
a 3 $\sigma$ peak is seen to the north-east 
of the protostar, which is likely caused by the differences 
of the emission peak position measured using 1 mm and 3 mm data. 
We estimate the protostar position by fitting a Gaussian to the highest 
resolution observations at 3 mm, and there is a slight 
offset relative to the protostar position measured using 1 mm data. 
If we shift the model with this offset at 1 mm, no residuals 
higher than 3 $\sigma$ are left in the small-scale image. 
The posterior-weighted results 
suggest a density power-law index $p \sim$ 2. 
In this case, the envelope density 
structure is similar to a singular isothermal sphere  
or the beginning stage of the Shu model 
with a very small infall region. 
In the Shu model, 
a free-fall-like $p \sim$ 1.5 profile is established quickly during the collapse process.
If the Shu model is applied strictly, an extremely young age of 
$\sim$10$^3$ yrs is implied.   
This age is much younger than other age estimates. For example, 
a kinematic age of $\sim$15,000 yrs is suggested by the outflow 
observations \citep{Bachiller2001}.    
The results are consistent with the single-wavelength 
study of \citet{Looney2003}, 
in which a larger sample of Class 0 YSOs are modeled and 
unphysical young ages are derived using the simple self-similar model.  
A steep density profile can be related to a finite mass reservoir, 
as the constraint from an outer boundary  
can steepen the density in the outer envelope  
\citep{Vorobyov2005c}. 
Another possibility is the change of dust grain properties 
across the envelope. 
If larger grains are present towards the inner envelope, their 
greater opacity can result in a steeper density profile 
being estimated.  
We do not investigate the radial variation of $\beta$ in our model, 
as no apparent $\beta$ variation is seen in Figure \ref{figBeta}.

\subsection{Spherical Power-law Envelope with an Inner Unresolved Component} \label{sec:m2} 

Disk formation is a natural consequence of angular momentum conservation 
when a rotating envelope collapses. 
It is expected to happen early in the star formation process, 
approximately in the Class 0 stage.
While characterizing disks in Class 0 YSOs, 
in particular their size and mass,   
is critical to reveal the mass accretion process, 
observing them is difficult due to the %
dusty envelopes around them. 
Distinguishing the disk component from the envelope emission 
requires a good understanding of the envelope,   
measuring the unresolved emission as the circumstellar disk component
\citep[e.g.,][]{Keene1990,Chandler1995}.

Although the pure power-law envelope can fit the observational 
data with statistical significance (\S \ref{sec:m1}), 
we add another parameter, a point source flux density, 
to the model to represent any unresolved component in our 
interferometric observations of L1157-mm. 
For simplicity, we assume that the dust properties for the 
unresolved component are the same as that for the rest of the envelope.  
Physically, this point source flux density is interpreted as an  
upper limit of the embedded disk component with a size smaller 
than the highest observational resolution of $\sim$0.3\arcsec~or 75 AU. 

\begin{figure}
\includegraphics[angle=0,width=1.0\textwidth]{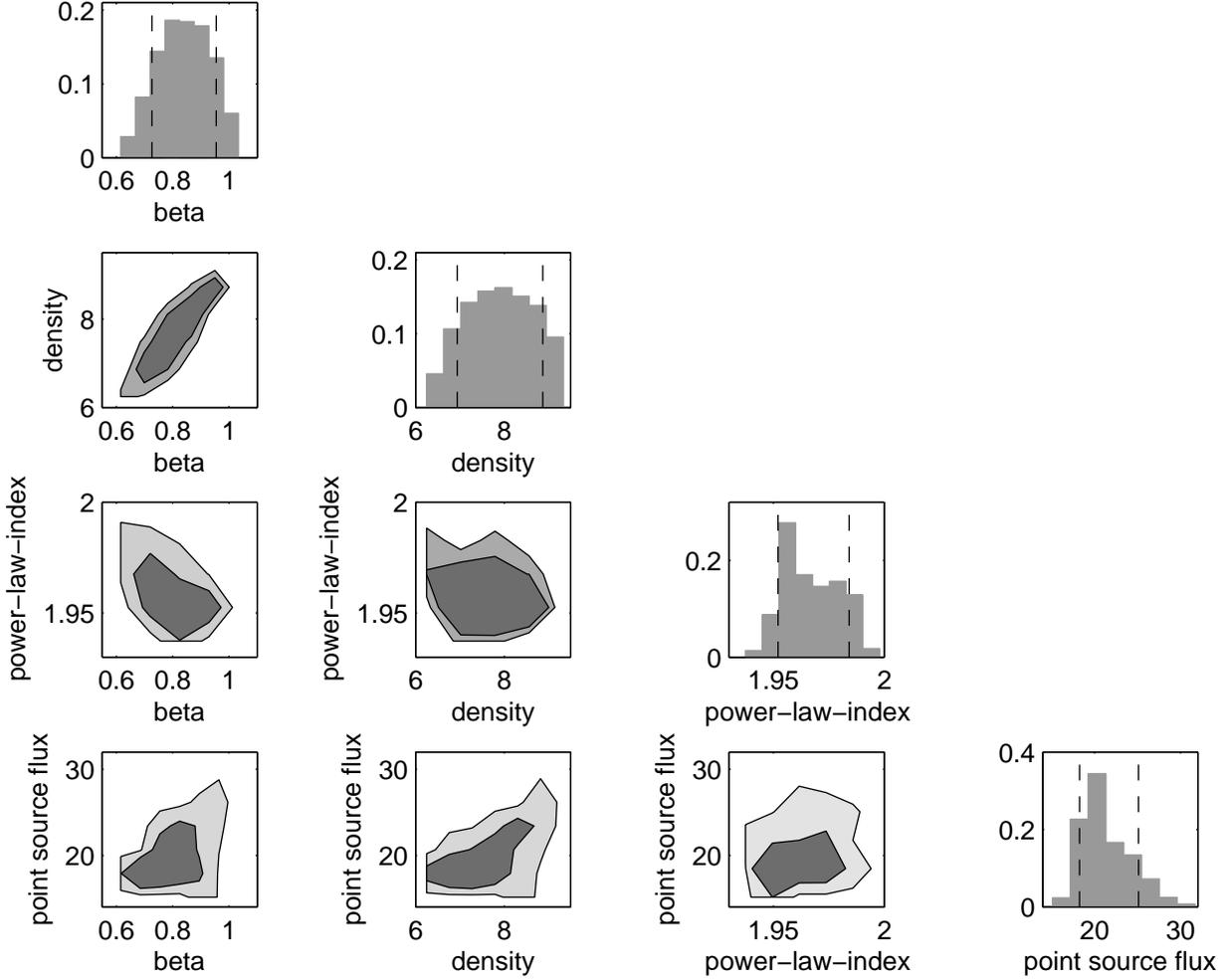}  
\caption[Marginalized posterior probability distributions of
the power-law model plus an unresolved component 
with absolute flux uncertainty. ]{ 
Same as Figure  \ref{figPDFm1_fs} for the spherical power-law model 
with an unresolved component. %
The absolute flux uncertainty is included. 
Marginalized posterior probability distributions for all four parameters
(dust opacity spectral index $\beta$, %
density $\rho_0$ at 100 AU in the unit of 10$^{-18}$~g~cm$^{-3}$, 
envelope density power-law index $p$, 
and point source flux density at 1 mm in the unit of mJy) 
are shown.  
In the
histograms, the dashed vertical lines enclose 68\% or 1 $\sigma$
confidence interval, with the expectation values and $\sigma$ listed in
Table \ref{tabM2}.  The dark and light areas in the 2-D contour plots
are the 68\% and 95\% confidence regions.
}
\label{figPDFm2_fs}
\end{figure}

\begin{figure}
\includegraphics[angle=270,width=0.95\textwidth]{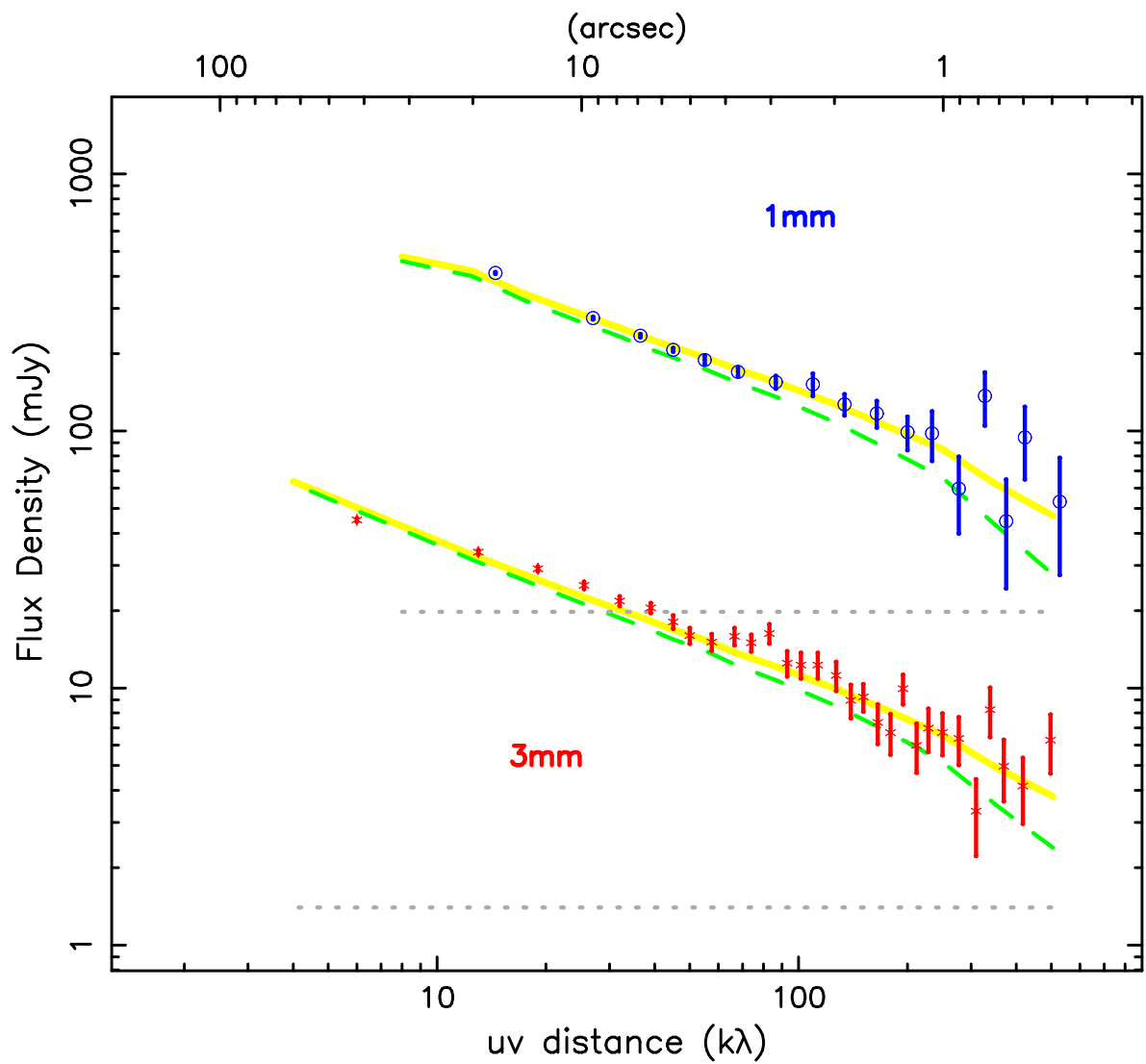} 
\caption[Visibility comparison of the power-law model plus an 
unresolved component. ]{ 
Same as Figure \ref{figBinM1} but for the power-law envelope model 
plus an unresolved component. 
Error bars are statistical errors in the annuli-bins only.  
Observational flux density, averaged vectorially and binned 
in $u$-$v$ annuli around the source center, are shown by 
circles for the 1 mm data and asterisks for the 3 mm data. 
The model fit with the marginalized parameters is shown 
by solid lines, which includes two components: 
a power-law envelope (broken lines) and 
an unresolved disk (dotted lines).  
}
\label{figBinM2}
\end{figure}

\begin{figure}
\includegraphics[angle=270,width=1.0\textwidth]{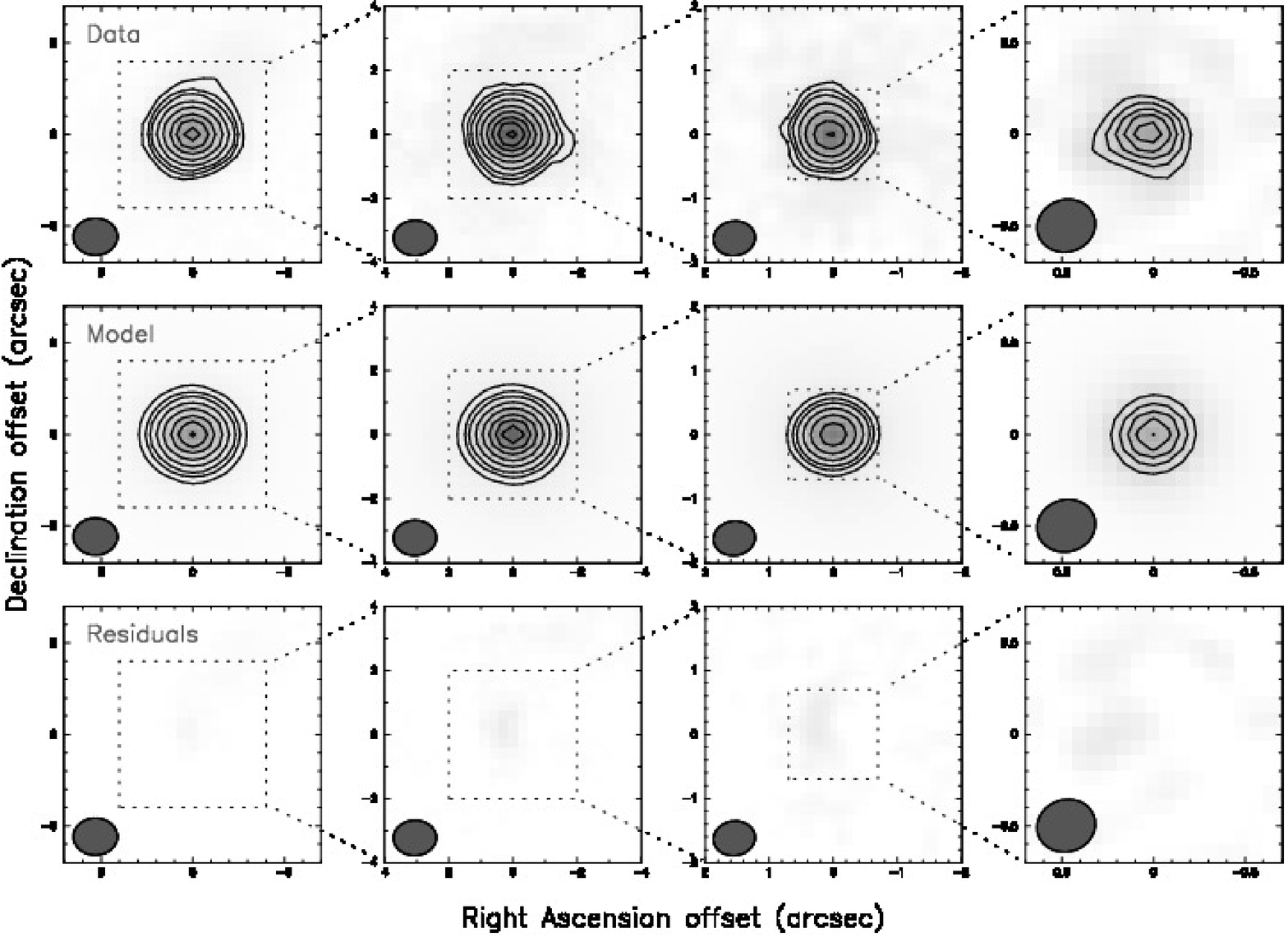}
\caption[Image comparison between data, model, and residuals 
at 3 mm. ]{ 
Same as Figure \ref{figImg3mmM1} but for 
the power-law envelope model with an unresolved component. 
Comparison between 3 mm dust continuum data ({\it upper row}), 
model ({\it middle row}), and residuals ({\it lower row}) 
of L1157-mm in the image space. 
 }
\label{figImg3mmM2}
\end{figure}

\begin{figure}
\includegraphics[angle=270,width=1.0\textwidth]{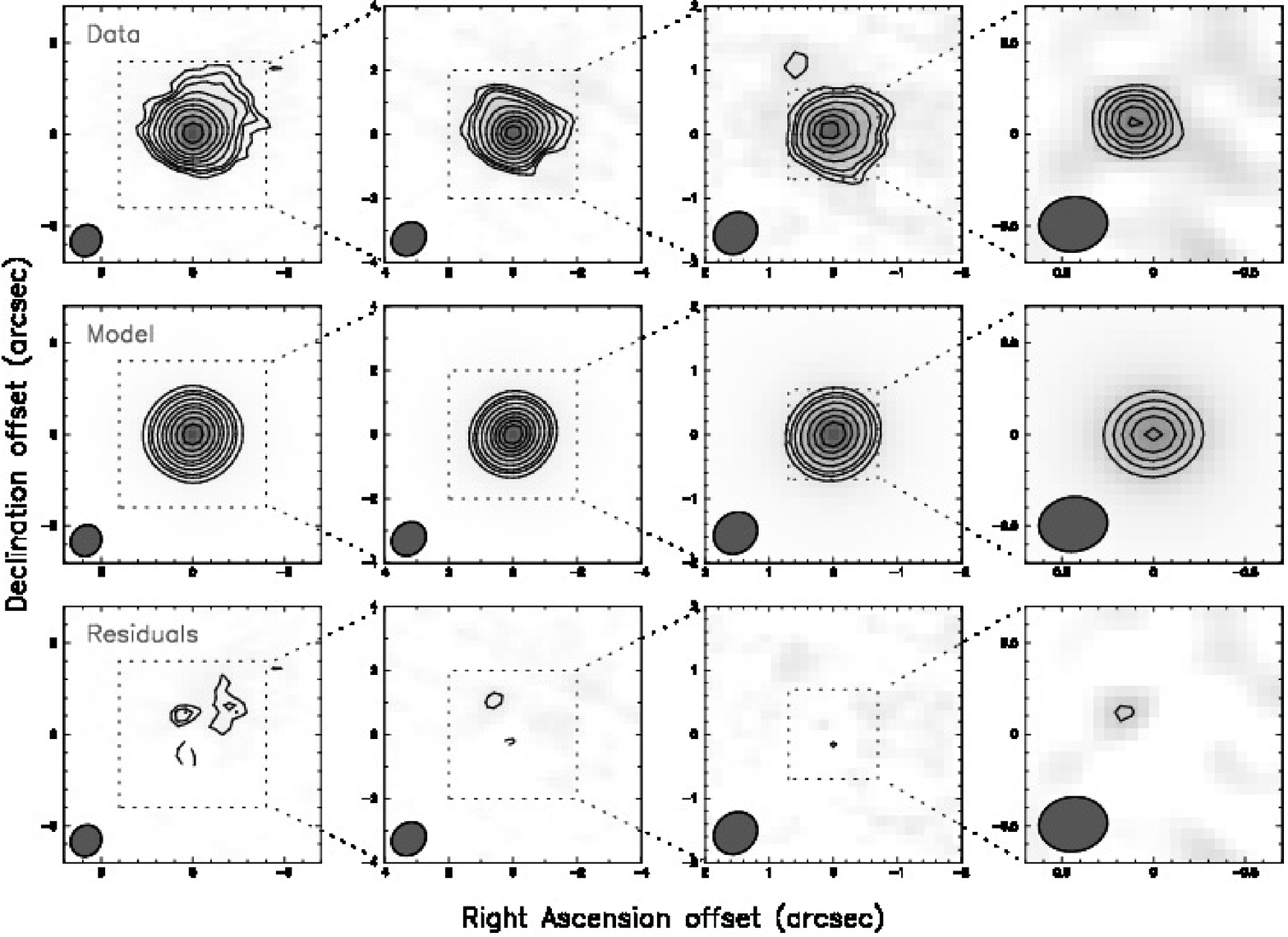}
\caption[Image comparison between data, model, and residuals
at 1 mm. ]{ 
Same as Figure \ref{figImg1mmM1} but for 
the power-law envelope model with an unresolved component. 
Comparison between 1 mm dust continuum data ({\it upper row}), 
model ({\it middle row}), and residuals ({\it lower row}) 
of L1157-mm in the image space. 
 }
\label{figImg1mmM2}
\end{figure}

Using the same technique as discussed in \S \ref{sec:m1}, 
we characterize the model parameters. 
Figure \ref{figPDFm2_fs} 
shows the result marginalized probability distributions, %
and Table \ref{tabM2} lists the expectation values and uncertainties.  
Visualization of the model-data comparison using the marginalized
parameters is shown in Figure \ref{figBinM2}, Figure \ref{figImg3mmM2},
and Figure \ref{figImg1mmM2} in the visibility domain and
the image domain.  
%
\begin{deluxetable}{lccc} 
\tabletypesize{\footnotesize} 
\tablecolumns{4} 
\tablewidth{0pc} 
\tablecaption{Power-law Envelope Model with an Unresolved Component} 
\tablehead{ 
\colhead{Parameter}  &  
\colhead{Mean }  &  
\multicolumn{2}{c}{Radius of the 68\% confidence interval} 
\\
\colhead{}  &  
\colhead{}  &  
\colhead{(statistical noise only)}  &  
\colhead{(with flux uncertainty)}   
\\
\colhead{(1)}  & 
\colhead{(2)}  & 
\colhead{(3)}  & 
\colhead{(4)}    
}
\startdata 
dust opacity spectral index $\beta$ \dotfill & 
0.84813  & 0.00038  & 0.11 
\\
%
density at 100 AU (in 10$^{-18}$~g~cm$^{-3}$)  \dotfill & 
 7.8627\tablenotemark{a} & 0.0053 & 0.96 
\\
density power-law index $p$  \dotfill & 
1.95004 & 0.00031 & 0.02
\\
unresolved 1 mm flux density (in mJy)  \dotfill & 
19.1835 & 0.0032 & 1.5 
\\
\enddata 
\tablenotetext{a}{
corresponding to a total envelope mass of 2.0662 M$_\odot$ 
}
\label{tabM2}
\end{deluxetable}

%
%
%
%
%
%
%
%
%
%
%
%
%
%
%
%
%
%
%
%
%
%
%


While a better fit with a smaller $\chi^2$ is achieved by the the power-law envelope model plus an unresolved component, 
the results %
are consistent with the results of the pure power-law envelope model 
presented in \S \ref{sec:m1}. 
The posterior-weighted mean of the density power-law index $p$ is 
slightly smaller than that of the pure power-law envelope model,  
due to the contribution of the point source flux.   
The dust opacity spectral index $\beta$ and density $\rho_0$ %
are consequently affected. 
With the added complexity of the model, 
larger uncertainties for the model parameters are obtained. 
In particular, 
the flux density of the unresolved component is %
not well constrained as it is not a necessary parameter. 
Nonetheless, the density index $p$ is still close to 2, so the inconsistency 
with the Shu model still exists (\S \ref{sec:m1}). 
The posterior-weighted mean of the unresolved flux is $\sim$2\% compared 
to the total flux measured by single-dish observations \citep{Gueth2003}, and 
$\sim$11\%  of the flux measured by CARMA.  
The flux density of the unresolved component can be converted to %
the upper limit of the embedded disk mass  
within the framework of the model. 
If we follow the empirical method of disk mass approximation in 
\citet{Looney2003} based on the disk modeling of HL Tau   
in \citet{Mundy1996},  
a disk of 0.05 M$_\odot$ at a distance of 140 pc 
is used as the standard candle for  
100 mJy emission at 2.7 mm. 
As a result, our marginalized model  
gives a disk mass of 4.1 M$_{Jup}$.
Alternatively, we can use a single-temperature optically thin source   
model to estimate the mass, that is,    
\begin{equation}
M =  F_\nu d^2 / \kappa_\nu B_\nu(T)  
\label{eqDiskMass}
\end{equation}
\citep{Hildebrand1983}. 
Following \citet{Looney2000}  
with the assumptions of T = 60 K and  
$\kappa$=0.1($\nu$/1200 GHz) cm$^2$ g$^{-1}$ (dust+gas),  
the estimated disk mass is 3.6 M$_{Jup}$.
Although these two methods of disk mass estimation give 
consistent results, 
the mass estimate is subject to the uncertainty of 
dust emissivity and temperature (see \S \ref{sec:disk}).

\subsection{Rotating Collapse Model} \label{sec:m3} 

Gravitational collapse of an envelope with uniform rotation 
has  been studied in \citet{Ulrich1976}, \citet{Cassen1981}, and 
\citet*[][hereafter the TSC model]{TSC1984}.  
The initial condition of the TSC model 
is a singular isothermal sphere, as in the Shu model.   
The non-zero angular momentum causes material to fall onto the midplane,  
following the streamline equation 
\begin{equation}
\frac{r}{r_c} = \frac{\sin^2\theta_0}{1-\cos\theta/\cos\theta_0} , 
\label{eqTSCstream}
\end{equation}
where $r_c$ is the centrifugal radius,  
$\theta$ is the angle from the rotation axis, %
and $\theta_0$ is the angle of the streamline at large $r$. %
A disk structure is expected inside the envelope with the density distribution  
\begin{equation}
 \rho = \frac{\dot M}{4 \pi (GMr^3)^{1/2}} 
( 1 + \frac{\cos\theta}{\cos\theta_0} )^{-1/2} 
( \frac{\cos\theta}{\cos\theta_0} + \frac{2 \cos^2\theta_0}{r/r_c} )^{-1}  . 
\label{eqTSC}
\end{equation}
 
We adopt the TSC model for the envelope fitting.   
The model parameters include: 
(a) the dust opacity spectral index $\beta$ as in Eq.~(\ref{eqKappa}),   
(b) the dust density $\rho_0$ at 100 AU, 
(c) the centrifugal radius $r_c$ of the TSC model 
in Eq.~(\ref{eqTSCstream}) and Eq.~(\ref{eqTSC}), 
and (d) a point source flux density (at 1 mm) 
to represent any unresolved component.  
The unresolved component is assumed to have the same dust properties 
as the envelope for scaling the 1 mm flux density to 3 mm. 
Besides that the TSC model implies an embedded disk, 
this point source flux density is required 
to obtain good fits  with the TSC envelope; 
we were not able to obtain a fit with 90\% confidence level 
with a zero unresolved flux. 

\begin{figure}
\includegraphics[angle=0,width=1.0\textwidth]{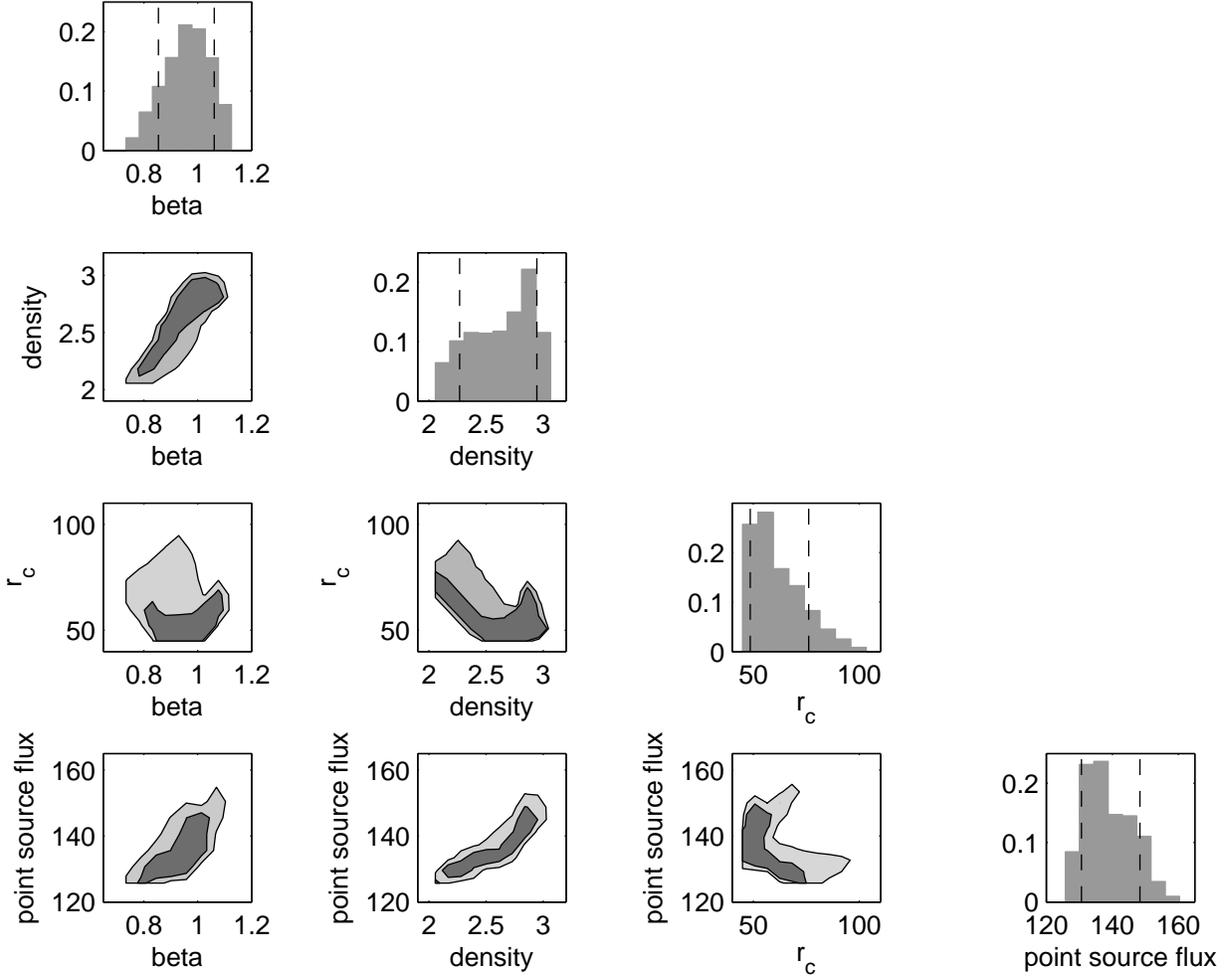}  
\caption[Marginalized posterior probability distributions of
the TSC model plus an unresolved component 
without absolute flux uncertainty. ]{ 
Same as Figure  \ref{figPDFm1_fs} but for the TSC model 
with an unresolved component. 
The absolute flux uncertainty is included. 
Marginalized posterior probability distributions for all four parameters
(dust opacity spectral index $\beta$, %
density $\rho_0$ at 100 AU in the unit of 10$^{-18}$~g~cm$^{-3}$, 
centrifugal radius $r_c$, and point source flux density at 1 mm in the unit of mJy) 
are shown.  
In the
histograms, the dashed vertical lines enclose 68\% or 1 $\sigma$
confidence interval, with the expectation values and $\sigma$ listed in
Table \ref{tabM3}.  The dark and light areas in the 2-D contour plots
are the 68\% and 95\% confidence regions.
}
\label{figPDFm3_fs}
\end{figure}

\begin{figure}
\includegraphics[angle=270,width=0.95\textwidth]{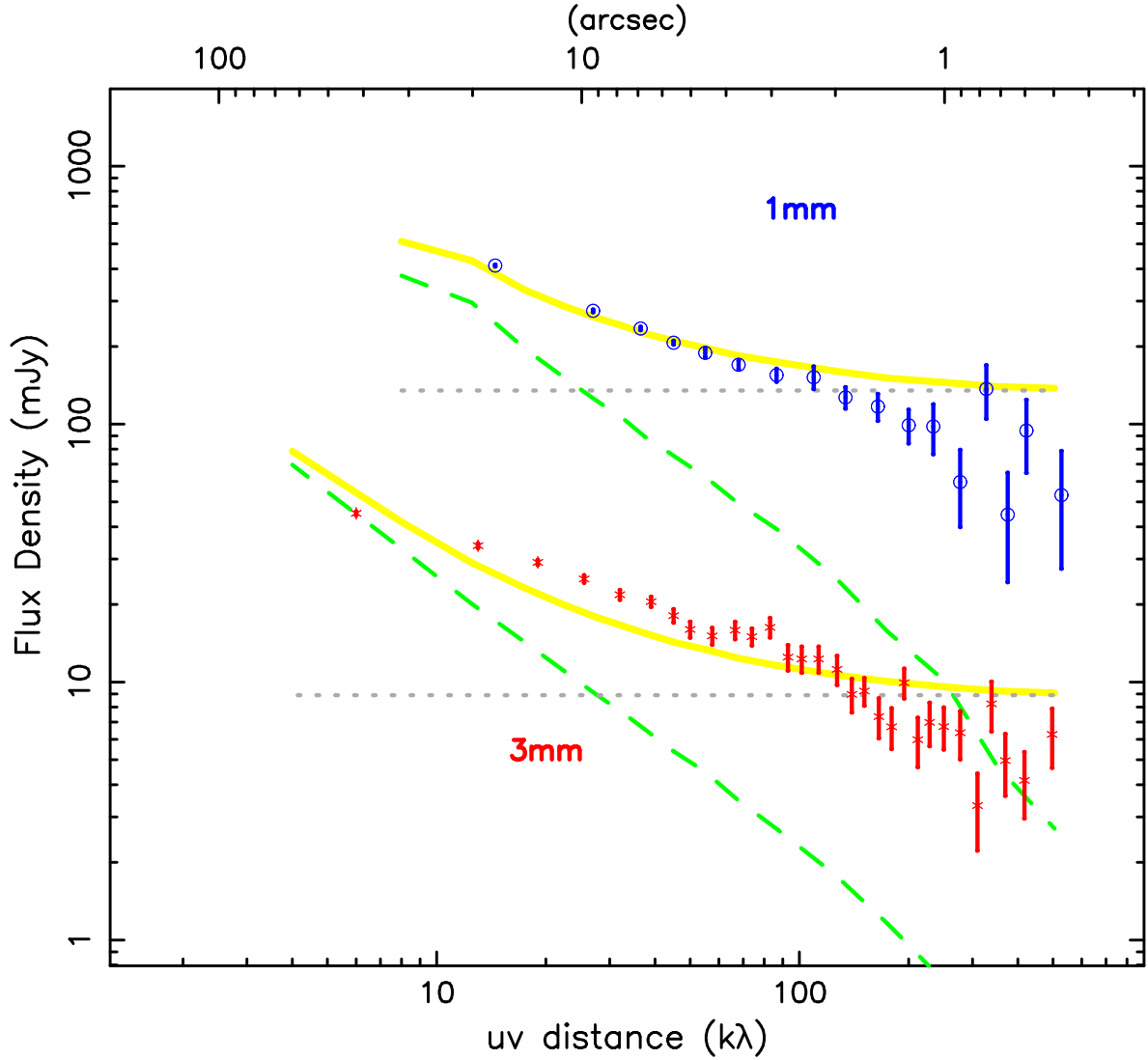} 
\caption[Visibility comparison of the TSC model plus an 
unresolved component. ]{ 
Same as Figure \ref{figBinM2} but for the TSC envelope model 
plus an unresolved component. 
The total model fit (solid lines) 
includes two components: 
an envelope (broken lines) and an unresolved disk (dotted lines).  
}
\label{figBinM3}
\end{figure}

\begin{figure}
\includegraphics[angle=270,width=1.0\textwidth]{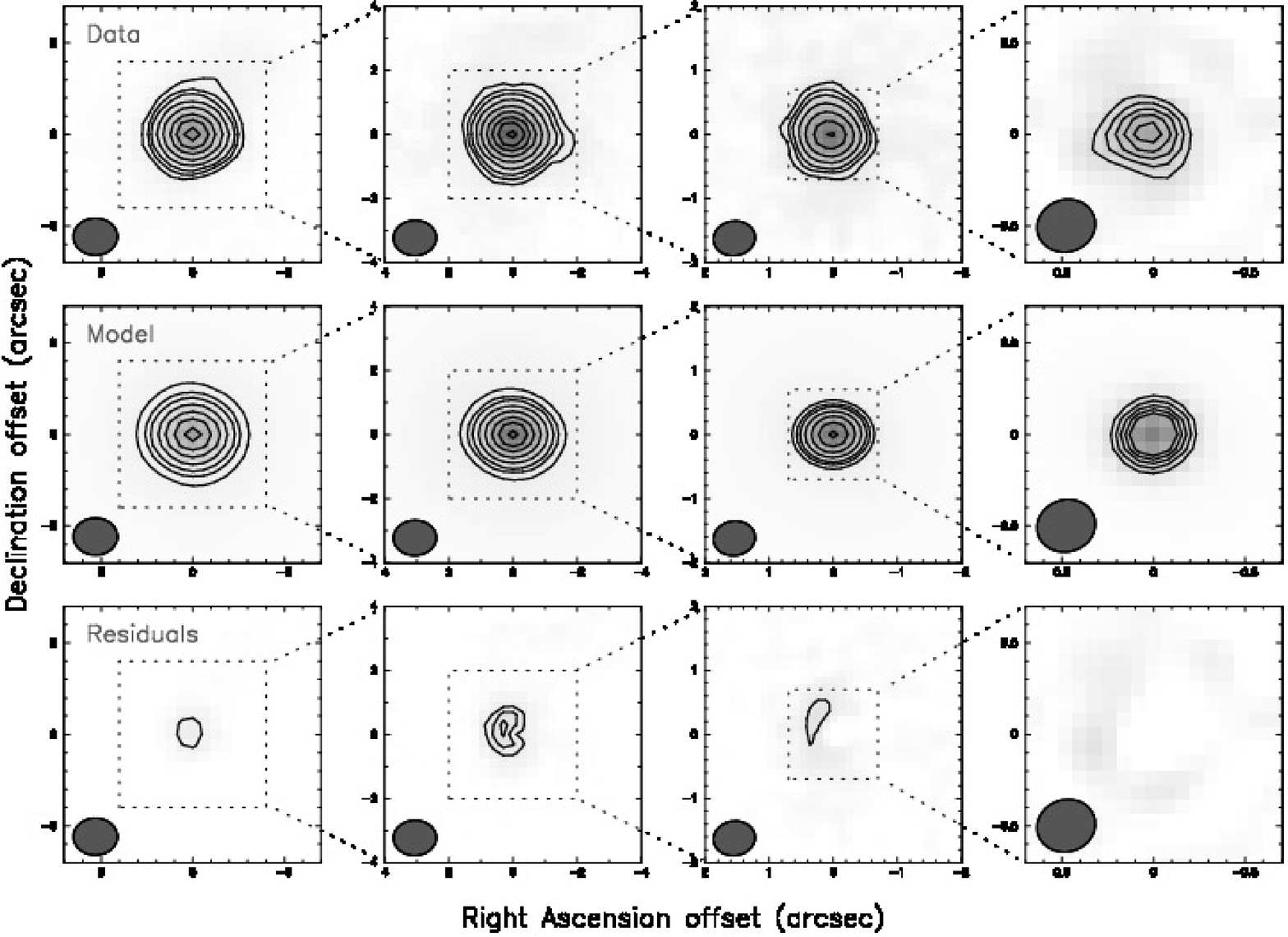}
\caption[Image comparison between data, model, and residuals 
at 3 mm. ]{ 
Same as Figure \ref{figImg3mmM1} but for 
the TSC model with an unresolved component. 
Comparison between 3 mm dust continuum data ({\it upper row}), 
model ({\it middle row}), and residuals ({\it lower row}) 
of L1157-mm in the image space. 
 }
\label{figImg3mmM3}
\end{figure}

\begin{figure}
\includegraphics[angle=270,width=1.0\textwidth]{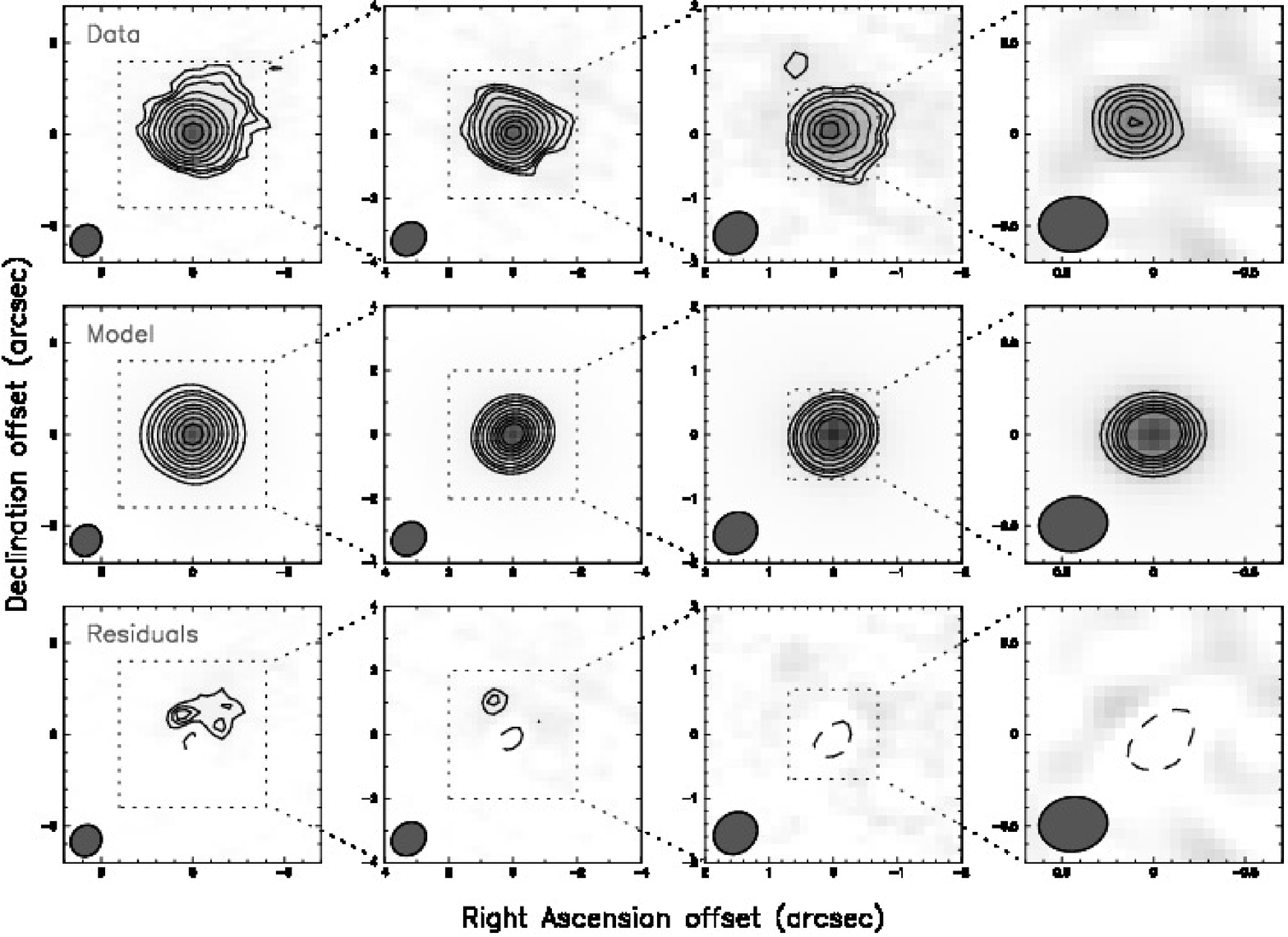}
\caption[Image comparison between data, model, and residuals
at 1 mm. ]{ 
Same as Figure \ref{figImg1mmM1} but for 
the TSC model with an unresolved component. 
Comparison between 1 mm dust continuum data ({\it upper row}), 
model ({\it middle row}), and residuals ({\it lower row}) 
of L1157-mm in the image space. 
 }
\label{figImg1mmM3}
\end{figure}

The model parameters are investigated 
as in \S \ref{sec:m1} and \S \ref{sec:m2}.  
Figure \ref{figPDFm3_fs} shows 
the marginalized probability distributions of model parameters,  
and  Table \ref{tabM3} lists   
the expectation values and uncertainties. %
Furthermore, 
visualization of the model-data comparison using the marginalized
parameters 
is shown in both the visibility domain and image domain  
in Figure \ref{figBinM3}, Figure \ref{figImg3mmM3}, 
and Figure \ref{figImg1mmM3}

\begin{deluxetable}{lccc} 
\tabletypesize{\footnotesize} 
\tablecolumns{4} 
\tablewidth{0pc} 
\tablecaption{TSC Model with an Unresolved Component} 
\tablehead{ 
\colhead{Parameter}  &  
\colhead{Mean }  &  
\multicolumn{2}{c}{Radius of the 68\% confidence interval} 
\\
\colhead{}  &  
\colhead{}  &  
\colhead{(statistical noise only)}  &  
\colhead{(with flux uncertainty)}   
\\
\colhead{(1)}  & 
\colhead{(2)}  & 
\colhead{(3)}  & 
\colhead{(4)}    
}
\startdata 
dust opacity spectral index $\beta$ \dotfill & 
0.96008    & 0.00022  & 0.10 
\\
density at 100 AU (in 10$^{-18}$~g~cm$^{-3}$)  \dotfill & 
2.7203\tablenotemark   & 0.0022 & 0.34 
\\
centrifugal radius $r_c$ (in AU)  \dotfill & 
44.9236    & 0.0016 & 14    
\\
unresolved 1 mm flux density (in mJy)  \dotfill & 
135.0764   & 0.0067 & 8.9   
\\
\enddata 
\tablenotetext{a}{
corresponding to a total envelope mass of 5.0266 M$_\odot$ 
}
\label{tabM3}
\end{deluxetable}


The TSC model only fits the data %
with a bright point source component. In this case, 
the envelope contributes little flux towards the total 
dust continuum, and 
the unresolved disk component dominates the emission at  
both 1 mm and 3 mm, especially at long baselines  
(see Figure  \ref{figBinM3}).  
While a rather shallow envelope model such as TSC is assumed, 
a strong small-scale component is required to fit the data. 
Similar results have also been seen in other studies   
\citep[e.g.,][]{Terebey1993,Enoch2009}. 
As discussed in \S \ref{sec:m2}, the unresolved flux density 
can be converted to the upper limit of disk mass.  
The posterior-weighted parameter implies an embedded disk of 
$\sim$25 M$_{Jup}$ 
using either method in \S \ref{sec:m2}.

Although the best-fit model passes a chi-square hypothesis test    
(or the null hypothesis is rejected with 90\% confidence),    
the TSC model does not fit the data as well as the power-law envelope model does. 
As seen in Figure \ref{figImg3mmM3} and Figure \ref{figImg1mmM3},   
the model does not subtract the data as cleanly as the power-law 
model does and leaves more residuals.  
In particular, residuals at 5 $\sigma$ level are seen 
in the 3 mm image (column 2 in Figure \ref{figImg3mmM3}).   
A worse fit is also shown by the larger $\chi^2$, which will be 
examined  in more detail in \S \ref{sec:modelSelection}.

\section{Discussion} \label{sec:discussion}

\subsection{Model Comparison} \label{sec:modelSelection}

A model is just a simplification of the unknown reality, 
but  we want to know 
which model provides a better approximation to all available data. 
In this section we apply model selection techniques 
and rank the models.  

By applying Bayesian inference at the model level,
one can use the Bayesian evidence for model selection. 
As discussed in Appendix B 
(Eq. (\ref{eqBayes}) and Eq. (\ref{eqBayesEvid})),  
the Bayesian evidence 
represents the probability of data given the model.  
It is marginalized over the full parameter space 
so the values of model parameters are not important, 
as opposed to parameter estimation 
for a particular model 
(\S \ref{sec:result}).  
%
%
For comparing two competing models, 
the ratio of evidence, also known as the Bayes factor,     
represents posterior odds and can infer whether one model is preferred 
over the other \citep[see ][for a review]{Liddle2009}. 

However, 
an exact method to compute the Bayesian evidence 
needs to fully evaluate likelihood in the entire parameter space 
and is very computationally expensive.  
The posterior distribution sampled by MCMC (\S \ref{sec:result})   
peaks around the maximum posterior probability, 
and is not sufficient to calculate the Bayesian evidence. 
Approximation such as the use of information-theoretic methods
is a good alternative approach for model selection  
\citep[e.g.,][]{Liddle2007}. 
For example, the 
Akaike information criterion \citep[AIC,][]{Akaike1974}, 
derived using  
the Kullback-Leibler information (or K-L distance), 
is defined as  
\begin{equation}  
\textrm{AIC} \equiv -2 \ln \mathcal{L}_{max} + 2 k   , 
\label{eq:AIC}
\end{equation}
where $\mathcal{L}_{max}$ is the maximum likelihood and 
$k$ is the number of model parameters. 
AIC provides a 
simple measure of how good the model approximates the 
information contained by the data, 
and a smaller value implies less information is lost and 
hence a better model. 
Detailed derivation and statistical implications can be found in 
\citet{BurnhamBook}. 
A second-order AIC, or AIC corrected, is suggested  
for small-sample bias adjustment as in  
\begin{equation}  
\textrm{AIC}_c = \textrm{AIC} + \frac{2k(k+1)}{N-k-1} ,  
\end{equation}
where $N$ is the number of data points. 
But in our case, $N \gg k$ so  AIC and AIC$_c$ converge.  
%
On the other hand, the  
Bayesian information criterion \citep[BIC,][]{BIC_Schwarz1978}, 
defined as 
\begin{equation}  
\textrm{BIC} \equiv -2 \ln \mathcal{L}_{max} + k \ln N ,  
\label{eq:BIC}
\end{equation}
is an approximation based on the Bayesian evidence ratio  
\citep[also see][]{Liddle2004}. 

A good model seeks for balance between goodness of fit and model simplicity.   
To obtain a better fit to the data or a smaller $\chi^2$, 
one may increase the model complexity with more parameters,   
but unnecessary use 
of parameters and over-fitting should be discouraged. 
The tradeoff is also seen in Eq. (\ref{eq:AIC}) 
and Eq. (\ref{eq:BIC}).  
As the best model minimizes AIC and BIC, 
smaller $\chi^2$ decreases the first term   
but extra parameters increase the second term. 
Model complexity in terms of the number of parameters 
is penalized in either AIC or BIC.

In this study, we evaluate AIC$_c$ and BIC for all models. 
Because either AIC$_c$ or BIC is on a relative scale, 
only the differences instead of actual values are meaningful 
\citep{BurnhamBook}. 
Results are listed in Table \ref{tabAICBIC}, where the model 
with the smallest value is preferred. 
As AIC$_c$ and BIC suggest different ranking between 
the pure power-law envelope model and 
the power-law envelope plus an unresolved component,   
we do not make an inference between them. 
However, the results show a large positive 
$\Delta$AIC$_c$ and $\Delta$BIC for the TSC model, implying that 
the power-law envelope model 
(either with or without the unresolved component) 
is decisively preferred against 
the TSC model plus an unresolved component.  
\begin{deluxetable}{lcccc} 
\tabletypesize{\scriptsize}
\tablecolumns{5} 
\tablewidth{0pc} 
\tablecaption{Model Comparison } 
\tablehead{ 
\colhead{Model }  &  
\colhead{\# of parameters}  &  
\colhead{$\chi^2_{min}$ = $\Delta$(-2 ln$\mathcal{L}_{max}$) \tablenotemark{a}}  &  
\colhead{$\Delta$AIC$_c$}  &  
\colhead{$\Delta$BIC}     
\\
\colhead{(1)}  & 
\colhead{(2)}  & 
\colhead{(3)}  & 
\colhead{(4)}  & 
\colhead{(5)}    
}
\startdata 
Power-law envelope \dotfill &
3 & 
9613473.1  &
1.9 & 0  
\\
Power-law envelope + an unresolved component \dotfill &
4 & 
9613469.2   & 
0  &  12.2 
\\
TSC envelope + an unresolved component \dotfill &
4 & 
9615698.7 & 
2229.5 & 2241.7
\\
\enddata 
\tablenotetext{a}{
The reduced $\chi^2_{min}$ are 0.9999793, 0.9999789, and 1.0000211.  
}
\label{tabAICBIC}
\end{deluxetable}



The model selection results are consistent with the 
{\it a posteriori} check presented in \S \ref{sec:result}.   
Compared to either the pure power-law envelope model 
or the power-law envelope plus an unresolved component, 
considerable residuals are seen  
in the image domain for the TSC model plus an unresolved component.

\subsection{Grain Growth} 

As mentioned in \S \ref{sec:dust}, dust properties are characterized 
by the opacity spectral index $\beta$ in our modeling.  
Depending on the environment, $\beta$ typically varies between 0 and 2. 
While $\beta \sim$ 2 implies small grains 
as in the interstellar medium, a smaller $\beta$ is usually 
found in many YSOs  
\citep[e.g.,][]{Natta2007PPV}. 
Decrease of $\beta$ can be caused by many factors, 
such as change of composition or grain geometry, 
but is usually associated with change of grain size distribution  
\citep[e.g.,][]{Krugel1994}. 
The $\beta$ value can be an evolutionary indicator of 
the dust grains in YSOs, 
as small grains in YSOs grow through coagulation, and eventually 
form  planets if conditions allow.  
Nevertheless, \citet{Miyake1993} studied the size effect and
showed that the observed decrease of $\beta$ in disk regions can be
explained by the growth of grain size without change of chemical composition.
If the grains in a dense region are composed of the same materials as in the
interstellar grains, the maximum grain size is expected to be larger than
3 mm to explain the observational results of
$\beta$ $\lesssim 1$ \citep{Draine2006}.
Grain growth has been the most widely accepted explanation for the
small $\beta$ found in protoplanetary disks
\citep[e.g.,][]{Beckwith2000PPIV,Draine2006}.

In the three envelope models presented in \S \ref{sec:result}, 
the posterior-weighted mean of $\beta$ range from 0.84 to 0.96.  
The absolute flux calibration dominates the uncertainty of $\beta$ 
estimation, resulting in a systematic uncertainty of $\sim$0.1.     
Still, $\beta$ being significantly smaller than the interstellar value is indicated. 
Our $\beta$ estimate for L1157-mm is %
in agreement with the samples of Class 0 YSOs in 
\citet{Jorgensen2007}, \citet{Kwon2009}, and \citet{Shirley2011m}.  
The result implies that dust grains in L1157-mm, and arguably most 
Class 0 YSOs, have gone through 
some grain growth to at least millimeter size 
from the initial interstellar grains. 
However, when exactly dust grains start to grow 
during the protostellar evolution is uncertain. 
For example, \citet{Ricci2010b} 
compared $\beta$ of YSOs with their evolutionary ages 
and did not find apparent trend or difference in $\beta$ 
for YSOs in different evolutionary stages.

As a uniform dust grain property is adopted in our simple dust model,  
our estimate of $\beta$ represents the grain property 
in the whole system, including disk and envelope.  
Depending on the assumed envelope model, 
the embedded disk can contribute a significant fraction of flux. 
Compared to grains in the envelope, 
grains in the disk are expected to be larger in size  
as an initial step of planet formation.  
Therefore, a smaller $\beta$ is expected in the disk than in the envelope.  
Besides, grain properties are likely to vary across the envelope,  
as has been observationally suggested for some Class 0 sources  
\citep[e.g.,][]{Chandler2000,Kwon2009}. 
Since our L1157-mm data are consistent with a single $\beta$ value 
accross the envelope, 
the radial dependence of $\beta$ or a disk with a different $\beta$ 
is not modeled in this study; 
to address the dust property change in YSOs 
requires a more complex model.

\subsection{The Earliest Circumstellar Disks } \label{sec:disk}

Circumstellar disks form as a physical consequence of 
angular momentum conservation 
when protostars accrete materials from their surrounding envelopes.
These planet-forming pre-main-sequence disks have been observed
and studied extensively 
\citep[e.g., see the reviews of][]{WilliamsCieza2011}.  
In particular, disk evolution from early Class 0 to Class I stage 
is interesting as it is the phase during which most mass accretion occurs. 
The mass and size of the disk %
grow rapidly as the system evolves and depend on the 
rotation rate %
and the magnetic field strength of the background cloud,   
as well as the detailed mechanisms of angular momentum redistribution 
\citep[e.g.,][]{TSC1984,Basu1998,Machida2011,Dapp2012}.

While the mass and size of these youngest disks are essential 
to reveal the early process of disk formation,  
observing them is, however, not straightforward. 
In addition to the limitations of observational resolution and
sensitivity, these disks are deeply embedded in their natal envelopes;  
probing them usually relies on indirect methods.
For example,
detection of water and methanol lines in Class 0 protostars 
can constrain the embedded circumstellar disks as the emission  
probably originates from the 
warm shocked layer of the disk-envelope interface, 
while different scenarios may not be ruled out  
\citep[e.g.,][]{Goldsmith1999L1157,Velusamy2002,Watson2007,Jorgensen2010}.   
Near-infrared scattered light images, showing a dark lane 
along the edge-on disk,   have also been used 
to infer the embedded disk structure \citep{Tobin2010b}.  
Another way to probe these youngest disks in embedded YSOs is 
through observations of dust continuum with a two-component model. 
The millimeter data consists of data continuum from both the disk and 
the envelope; 
with accurate modeling of the envelope, the circumstellar disk 
comopnent can be separated. 
In other words, the disk component is measured as the residual
emission with the envelope contribution subtracted 
\citep[e.g.,][]{Keene1990,Looney2003,Harvey2003a,Jorgensen2005,Chiang2008,Enoch2009}.
This method avoids the complexity of chemical effects, but 
requires the use of a theoretical envelope model. 
Using a two-component model, attributing the entire unresolved 
flux to be from the embedded disk, and following   
the mass estimation in \citet{Looney2000} and \citet{Looney2003},  
we derive the disk mass to be 
$\sim$4 M$_{Jup}$ assuming a power-law envelope,  
or $\sim$25 M$_{Jup}$ assuming a TSC envelope. 
Note that this mass estimate is an upper limit valid only 
within the framework of the models, and 
is highly dependent of 
the assumed envelope structure, dust opacity,  
disk temperature, and optical depth.  
For the envelope structure, we have shown that the 
power-law envelope model is preferred against the TSC model. 
But a large uncertainty still exists. 
For example, if instead we adopt a different dust opacity   
$\kappa$=0.015($\nu$/300 GHz) cm$^2$ g$^{-1}$ (dust+gas) 
and a single temperature of 30 K, as used in \citet{Greaves2011},    
the disk mass is 
$\sim$13 M$_{Jup}$ assuming the power-law envelope model.  
Our deduced disk mass is low compared to the estimate 
in \citet{Greaves2011}, where a disk mass of 80 M$_{Jup}$ is obtained for L1157. 
Although Eq.~(\ref{eqDiskMass}) is also used in \citet{Greaves2011},  
they assume the compact emission at 2\arcsec~is solely from the disk 
and no envelope subtraction is done, 
which possibly causes the differences in the estimated disk mass.

On the other hand, 
an empirical method that  
measures the small-scale flux at baseline $\sim$50 k$\lambda$
has also been used to estimate the unresolved disk component in embedded YSOs   
\citep{Jorgensen2009,Enoch2011}. 
This is based on the presumption %
that the envelope contributes little flux $\gtrsim$50 k$\lambda$.   
The 1 mm flux density of L1157-mm is around 200 mJy at 50 k$\lambda$, 
implying a disk mass of $\sim$174 M$_{Jup}$  
using $\kappa_{1.3mm}$=0.009 cm$^2$ g$^{-1}$ \citep{Enoch2011}. 
The estimated disk mass is comparable to other Class 0 
sources  in \citet{Enoch2011} and lighter than those 
in \citet{Jorgensen2009}, but much heavier than our disk mass 
estimate using a two-component model.   
The empirical method of \citet{Jorgensen2009} and \citet{Enoch2011} 
seems to over-estimate the disk component in L1157-mm;
one reason is that the flux density 
drops significantly longward of 50 k$\lambda$ 
(Figure \ref{figConfigs}) 
so the flux from the unresolved disk is 
apparently lower than 200 mJy.  
In addition, a relatively shallow envelope profile 
is assumed in either study 
(power-law of 1.5 in \citealt{Jorgensen2009} 
and TSC in \citealt{Enoch2011}), and the assumed envelope 
structure considerably affect the flux ratio from envelope 
and disk.  %
Since our observations do not resolve %
the circumstellar disk, we can put an upper limit on 
the disk size for L1157-mm.  
Assuming a distance of 250 pc, the upper limit of 
the disk radius is $\sim$40 AU. 
This is consistent with the theoretical scenario of \citet{Dapp2012} in which 
Class 0 disks are small.  
This size constraint for L1157-mm is also consistent with 
those in other Class 0 YSOs. 
For example, 
no disks are detected 
for a sample of 5 Class 0 objects in \citet{Maury2010} with a resolution down to 
0.3\arcsec, corresponding to a size scale of 42-75 AU, and   
the embedded disk at the edge-on Class 0 YSO VLA 1623A 
is constrained to be smaller than 50 AU in radius    
\citep{Ward-Thompson2011}.  
On the contrary, the disk embedded in the edge-on Class 0 YSO 
L1527 has been resolved by 7 mm VLA observations in  
\citet{Loinard2002}; 
the disk structure is also seen with  
SMA and CARMA observations (Tobin et al. 2012).  
The size of Class 0 disks is comparable to or smaller than 
the size of older circumstellar disks 
\citep[e.g.,][]{Eisner2005,Andrews2009,Vicente2005}, 
but no clear trend can be inferred at this point. 

\section{Summary } \label{sec:summary} 

\begin{enumerate}
\item
Multi-configuration CARMA observations 
of the edge-on Class 0 YSO L1157-mm are presented. 
In our dust continuum data at both 1 mm and 3 mm, 
a nearly spherical circumstellar envelope is seen at the 
size scale of $\sim$10$^2$ to $\sim$10$^3$ AU. 
No circumstellar disk on the small scale is resolved. 

\item
Radiative transfer modeling is performed to compare the interferometric data 
with the theoretical envelope models.  
A power-law envelope and a TSC envelope are considered. 
We add an unresolved component 
to represent the embedded disk. 
Bayesian inference is employed for parameter estimation. 
The absolute amplitude uncertainty, resulting from the flux calibration of 
the data reduction process, plays a critical role in parameter errors. 

\item
A density index $p \sim$ 2 is suggested for the power-law 
envelope, consistent with the results in \citet{Looney2003} 
for a larger sample of Class 0 YSOs.   
An unphysical young age is suggested if the Shu model is applied strictly. 
The data can be fitted by a pure power-law envelope without a compact emission 
from the embedded disk component.

\item
The dust grain properties of the envelope are studied through the 
dust opacity spectral index $\beta$.  The result 
$\beta \sim$ 0.9 is significantly smaller than the $\beta$ value 
in the interstellar medium, implying that 
grain growth has already started in L1157-mm. 

\item
The unresolved disk component is constrained    
to be $\lesssim$40 AU in radius and 
$\lesssim$4-25 M$_{Jup}$ in mass. 
However, the mass estimate of the embedded disk component 
heavily relies on the assumed envelope model 
as well as the assumed disk characteristics.  
%
For example, 
a shallow envelope, such as the TSC model with a density power-law index $p
\sim$ 1.5 in the outer region, requires a strong point source flux
from the unresolved disk, while a steep envelope with $p \sim$ 2
can fit the observational data without an embedded disk.

\item
Different envelope models are compared using 
an information-theoretic approach. 
The results prefer the power-law envelope model 
against the TSC model, which is also shown in the 
{\it a posteriori} check in the image domain.

\item
This is the first study that utilizes 
the Bayesian techniques and model selection to consider multiple 
envelope models and make statistical inference for embedded YSOs.  
Future observations, especially high-resolution ALMA observations, 
will resolve the transition zone between the envelope and the disk,  
and further constrain the structures of Class 0 YSOs.

\end{enumerate}

\acknowledgements{

The authors thank B. Reipurth and the anonymous referee for their 
careful reading of the manuscript and helpful comments.  
H.-F.C. and  L.W.L. acknowledge support from the Laboratory for Astronomical 
Imaging at the University of Illinois and the NSF under grant AST-07-09206. 
J.T acknowledges support provided by NASA through Hubble Fellowship grant \#HF-51300.01 awarded by the Space Telescope Science Institute, which is operated by the Association of Universities for Research in Astronomy, Inc., for NASA, under contract NAS 5-26555. 
We also thank CARMA staff and observers for their assistance in obtaining the data.  
Support for CARMA construction was derived from the states of Illinois, California, and Maryland, the James S. McDonnell Foundation, the Gordon and Betty Moore Foundation, the Kenneth T. and Eileen L. Norris Foundation, the University of Chicago, the Associates of the California Institute of Technology, and the National Science Foundation. Ongoing CARMA development and operations are supported by the National Science Foundation under a cooperative agreement, and by the CARMA partner universities. 

}

{\it Facilities:} \facility{CARMA ()}

\appendix 
\section{Error Estimate of the Approximate Dust Opacity Spectral Index}


In the optically thin limit and the Rayleigh-Jeans regime, 
the dust opacity spectral index  $\beta_{thin}$ can be approximated 
using the flux density at two wavelengths. 
In this appendix we discuss the error propagation from the 
observational uncertainty to the deduced $\beta_{thin}$ value. 
Let $F_1$ and $F_2$ be the flux density 
at frequencies $\nu_1$ and $\nu_2$, $\beta_{thin}$ can be expressed 
as in Eq.~(\ref{eqBeta}): 
\begin{equation} 
\beta_{thin} = \frac{\ln F_1 - \ln F_2}{\ln \nu_1 - \ln \nu_2} - 2.  
\label{eqBetaT}
\end{equation} 
Assuming $F_1$ and $F_2$ are indenpendent variables 
with standard deviations $\sigma_{1}$ and $\sigma_{2}$, 
the standard error propagation gives 
\begin{equation} 
\sigma_{\beta_{thin}}^2 = 
  \left| \frac{\partial \beta_{thin}}{\partial F_1} \right| ^2 \sigma_{1}^2 
 + \left| \frac{\partial \beta_{thin}}{\partial F_2} \right| ^2 \sigma_{2}^2 .
\label{eqError} 
\end{equation} 
Taking the partial derivative of Eq.~(\ref{eqBetaT}), we obtain 
\begin{equation} 
\frac{\partial \beta_{thin}}{\partial F_1} = \frac{1}{(\ln \nu_1 - \ln \nu_2) F_1} 
\label{eqPar1}
\end{equation} 
and 
\begin{equation} 
\frac{\partial \beta_{thin}}{\partial F_2} = -\frac{1}{(\ln \nu_1 - \ln \nu_2) F_2}. 
\label{eqPar2}
\end{equation} 
Replacing  Eq.~(\ref{eqError}) using Eq.~(\ref{eqPar1}) and Eq.~(\ref{eqPar2}), 
the uncertainty of the derived  $\beta_{thin}$ is then 
\begin{equation} 
\sigma_{\beta_{thin}}^2 =  \left(\frac{1}{\ln \nu_1 - \ln \nu_2}\right)^2 
              \left(   \frac{\sigma_{1}^2}{F_1^2} 
                    +  \frac{\sigma_{2}^2}{F_2^2}  \right)  .
\label{eqBetaErr} 
\end{equation} 

Using Eq.~(\ref{eqBetaErr}) and 
assuming the absolute flux error can be represented as a Gaussian 
noise with a standard deviation of 10\%, that is, 
$\sigma_{1} = 0.1 F_1$ and  $\sigma_{2} = 0.1 F_2$, 
the uncertainty of $\beta_{thin}$ is $\sim$0.15 for our data at 229 and 91 GHz.   
Note that here the error is assumed be normally distributed, 
different from the flat prior assumed in \S 4. 



\section{Fitting Technique and Statistical Inference} 

%
In this appendix we give a brief introduction to Bayesian inference,  
as opposed to frequentist statistics. 
Also, we describe our technical procedure to characterize model parameters 
and their uncertainty. 

The main concept of Bayesian inference is to 
incorporate prior knowledge on the hypothesis.   
Also, information is represented in terms of 
a probability density function (PDF) in parameter space. 
Mathematically, given the observed data, 
the posterior probability of model parameters 
can be specified by the Bayes' theorem
\begin{equation}
 P(x|D,M) = \frac{P(D|x,M)P(x|M)}{P(D|M)}  
\label{eqBayes}
\end{equation}
where $x$ stands for model parameters, $D$ stands for data, 
$M$ denotes a particular model with its model assumptions 
and other background information,  
$P(x|M)$ is the prior probability of model parameters $x$, 
$P(D|x,M)$ is the conditional probability or 
the likelihood of data given the model with parameter $x$, 
and $P(D|M)$ is the evidence or global likelihood. 
The evidence $P(D|M)$ is the net probability of the data given the model, 
as it sums the product of likelihood and prior over parameter space: 
\begin{equation}
P(D|M) = \int P(D|x,M) P(x|M) d x .  
\label{eqBayesEvid}
\end{equation}
The evidence is independent of the parameter values 
and can be seen as a normalizing factor in Eq. (\ref{eqBayes}); 
therefore it is not important for parameter estimation of a single model. 
For the same reason, the model label $M$ is sometimes omitted when 
only one model is considered. 
However, evidence is useful for comparing multiple models (\S \ref{sec:modelSelection}). 

Whether to view a statistics problem with Bayesian or frequentist 
approach is under debate. 
The disputes are beyond the scope of this study   
and more discussions can be found in \citet{Loredo1990,Loredo1992}.  
Nevertheless, in the case of a uniform prior, 
the method of using the posterior probability is equivalent to  
maximum likelihood estimate 
as far as identifying the best-fit parameter values is concerned. 
As for full parameter estimation including estimating the parameter errors, 
different approaches are adopted for Bayesians 
and frequentists, and will be discussed later.  

The likelihood of data given the model is 
characterized by $\chi^2$ (as defined in Eq. (\ref{eq:chi2})) and  
$ P(D|x) \sim$~exp$(-\chi^2(x,D)/2) $ 
with model parameter $x$ and observational data $D$.  
Therefore the most probable parameters with the maximum likelihood  
can be obtained by locating the global minimum of $\chi^2$.  
We use the Nelder-Mead simplex algorithm as implemented in MATLAB 
with bound constraints to search for the minimum.   
Besides fast convergence, 
this method does not evaluate function derivative, which suits our application 
because fewer modeling evaluations are required.  
Several starting points are used to look for several convergent minimums, 
and they are checked to be consistent with each other.  
This is to make sure that what is found is the global minimum, not a local minimum.

Once the best-fit parameter values are identified, 
efforts are made to characterize the uncertainty.  
The essence of parameter estimation is to 
characterize the reliability of an estimate on model parameters 
under the assumption that the best-fit model is correct. 
In other words, the best-fit model needs to 
show statistical significance  %
based on a hypothesis test %
before any of the following parameter estimation can make sense. 
For example, in the standard Pearson's chi-square hypothesis test, 
$\chi^2$ value of the model needs to be smaller than a critical value 
depending on the degrees of freedom to reject the null hypothesis. 
In the following we discuss two common methods of parameter estimation: 
(1) a frequentist approach to characterize $\Delta \chi^2$   
and infer statistical significance, 
and (2) Markov chain Monte Carlo (MCMC) in the context of Bayesian inference.

The frequentist $\Delta \chi^2$ statistics has been suggested in 
Lampton et al. (1976) and Avni (1976), and summarized in Press et al. (2002).     
With the definition of $\Delta \chi^2(x) = \chi^2(x) - \chi^2(x_{best\mbox{-}fit}) $,  
$\Delta \chi^2(x)$ is chi-square distributed with $p$ degrees of freedom,  
where $p$ is the number of fitted parameters or parameters of interest.  
Then the level of confidence can be estimated according to 
the chi-square distribution.   
Although it relies on the validity of the best-fit model,  
the exact value of $\chi^2(x_{best\mbox{-}fit})$ is not important for parameter 
estimation. 
The $\Delta \chi^2(x)$ statistics is independent of the Pearson's chi-square test and $\chi^2(x)$ statistics,  
and focuses on the variation of $\chi^2(x)$ in parameter space $x$.   
A common way to illustrate the results is through 
iso-chi-square contours %
or hyper-surface %
in multi-dimensional parameter space as the confidence region. 
In the case that only partial parameters are of interest, 
the remaining nuisance parameters should be varied to minimize $\Delta \chi^2(x)$ 
instead of direct projections. 
(c.f. In Bayesian inference, the nuisance parameters are marginalized over.) 
For example, when only one parameter is of interest, 
$\Delta \chi^2(x)$ is distributed as a chi-square distribution 
with one degree of freedom.  
The 68\% confidence interval corresponds to  
the region bounded by $\Delta \chi^2(x) = 1$.

Despite the controversy over the flaws of applying this method with nonlinear models 
\citep{Loredo1992}, estimating $\Delta \chi^2(x)$ over a large parameter space 
can be computationally difficult.  
A grid on parameters or equivalent technique is required. 
The large number of evaluations usually makes this method impractical, 
especially when the number of parameters is large \citep{Ford2005}.

On the other hand, MCMC offers a  
very efficient way to estimate the posterior probability 
in Bayesian inference, 
compared to any other methods that require grid searching. 
Rather than minimizing on each grid point and probing the 
variation of $\chi^2(x)$, 
the posterior probability respect to parameters of interest 
is estimated through marginalization over all other parameters.  
For example, given a PDF $P(x_1,x_2|D)$ where 
$x_1$ is the parameter of interest 
and $x_2$ is a nuisance parameter, %
the $x_2$ space is integrated over according to probability 
to obtain the marginalized PDF, as in  
\begin{equation}
P(x_1|D) = \int P(x_1,x_2|D) d x_2 = \int P(x_1|x_2,D) P(x_2|D) d x_2 .   
\end{equation}
At first glance, a straightforward marginalization can be 
very computationally extensive, similar to the necessity of a grid evaluation 
in frequentist methods. 
However, marginalized results can be obtained efficiently with MCMC.  
One of the reasons is that it searches the parameter space according to probability   
and the parameter space with low probability is less explored 
and sometimes not probed at all.

The Metropolis-Hastings algorithm of the MCMC method is utilized to 
construct the Markov chain. 
Markov chain is a sequence of parameter values representing the system 
and characterized by a transition probability that controls the  
random process from one state to another.   
The transition probability and the next state 
are only dependent of the current state, but not any previous states.   
Regardless of the starting state, the chain eventually  
converges to a stationary or equilibrium distribution according the PDF.  
We use the Metropolis-Hastings algorithm to draw the sample and construct the chain. 
This algorithm uses a proposal distribution $q(x'|x)$, 
or the candidate transition probability distribution function,  
to generate a trial state $x'$ based on the current state $x$. 
Then the proposed state is randomly accepted with the acceptance probability 
\begin{equation}
 \alpha(x'|x) = \min\left[\frac{P(x'|D)q(x|x')}{P(x|D)q(x'|x)} ,1\right], 
\end{equation}
or otherwise rejected.   
The arrangement results in a 
transition probability 
\begin{equation}
 T(x'|x) = q(x'|x) \alpha(x'|x)   
\end{equation}
which is 
reversible ($\pi(x) T(x'|x) = \pi(x') T(x|x')$, 
where $\pi(x)$ is the equilibrium probability at state $x$)  
and irreducible (possible to go from any state to any state). 
As introduced earlier, $P(x|D)$ is the posterior PDF 
given the observational data, 
and approximately proportional to exp$(-\chi^2(x,D)/2)$ with flat prior.  
Practically, $\chi^2(x,D)$ is evaluated at each proposed state change.  
 
The Metropolis-Hastings algorithm assures that 
the chain converges to $P(x|D)$ as the sample number is large. 
The convergence rate is related to the choice of $q(x'|x)$.  
A typical choice is a Gaussian function centered around x, that is, 
\begin{equation}
 q(x'|x) = \frac{1}{\sqrt{2\pi w^2}} \exp\left(-\frac{(x'-x)^2}{2 w^2}\right) = q(x|x') . 
\end{equation}
The width of the Gaussian, specified by $w$, determines 
the trial step size.   If the step size is too large, most 
trial states are rejected so the calculation becomes very inefficient; 
if the step size is too small, the chain behaves like a random walk 
and requires a long time to converge.  
An optimal acceptance rate is suggested to be around 0.23 
for multi-dimensional parameter space \citep{GelmanBook}.  
The choice of a symmetric proposal distribution also reduces the 
acceptance probability into a simper form 
\begin{equation}
 \alpha(x'|x) = \min\left[\frac{P(x'|D)}{P(x|D)} ,1\right]  
         = \min\left[\exp\left(\frac{\chi^2(x,D)-\chi^2(x',D)}{2}\right),1\right].  
\end{equation}

The posterior probability is obtained with a converged Markov chain, 
as its density of points in parameter space follows 
the posterior probability of the parameters.  
Marginalization is done through projecting the Markov chain 
to the space of parameters of interest. 
We estimate the 68\% and 95\% confidence limits      
and the corresponding standard deviation %
based on the simulated MCMC. 
Specifically, the 68\% or 1 $\sigma$ confidence limit encloses  
68\% of  accepted points along the Markov chain, and 
represents the region containing 68\% of 
the total probability distribution.  
We also report the expectation value of each parameter, 
weighted by the posterior marginalized probability as in 
\begin{equation}  %
\langle x_i \rangle = \int P(x_{i}|D) x_{i} d x_{i} 
      = \frac{1}{N} \Sigma_j x_{i,j}
\end{equation}
where $j$ denotes the points in the Markov chain and 
$N$ is the total number of points.  
The expectation values from the marginalized distributions  do not 
need to be identical to the parameters with the maximum likelihood,  
because we are projecting the values from a high dimensional 
distribution which may not be a multivariate Gaussian; 
however, they should be consistent.

\bibliographystyle{apj}
\bibliography{ref}
\end{document}